\documentclass[preprint,preprintnumbers,amsmath,amssymb]{revtex4}
\usepackage{graphicx}
\usepackage{dcolumn}
\usepackage{bm}
\usepackage{verbatim}
\usepackage{xcolor}
\usepackage{epsfig}
\usepackage{bbm}
\usepackage{epstopdf}
\usepackage{hyperref}
\usepackage{amsmath,mathtools}
\usepackage{natbib}
\usepackage[caption=false]{subfig}
\usepackage{dsfont}
\usepackage{multirow}
\usepackage{physics}
\usepackage{cellspace}
\newcommand\beq{\begin{equation}}
\newcommand\eeq{\end{equation}}
\newcommand\bea{\begin{eqnarray}}
\newcommand\eea{\end{eqnarray}}

\DeclareMathOperator{\erfi}{erfi}

\def\sgn{{\rm sgn}}
\usepackage{xcolor}

\def\Tr{{\rm Tr}}

\newcommand\LG[1]{\rm LG{#1}}

\def\x0{{{\bf x}_0}}

\begin{document}
	
	
	\title{Leggett-Garg tests for macrorealism in the quantum harmonic oscillator and more general bound systems}

	\author{C. Mawby}
	\email{c.mawby18@imperial.ac.uk}
	\author{J.J.Halliwell}%
	\email{j.halliwell@imperial.ac.uk}
	\numberwithin{equation}{section}
	\renewcommand\thesection{\arabic{section}}

	\affiliation{Blackett Laboratory \\ Imperial College \\ London SW7
		2BZ \\ UK }

	
	\begin{abstract}
The Leggett-Garg (LG) inequalities were introduced to test for the possible presence of macroscopic quantum coherence. Since such effects may be found in various types of macroscopic oscillators, we consider the application of the LG approach to the one-dimensional quantum harmonic oscillator and also to more general bound systems, using a single dichotomic variable $Q$ given by the sign of the oscillator position.  We present a simple method to calculate the temporal correlation functions appearing in the LG inequalities for any bound system for which the eigenspectrum is (exactly or numerically) known. We apply this result to the quantum harmonic oscillator for a variety of experimentally accessible states, namely energy eigenstates, and superpositions thereof. For the subspace of states spanned by only the ground state and first excited state, we readily find substantial regions of parameter space in which the LG inequalities at two, three and four times can each be independently violated or satisfied. We also find that the violations persist, although are reduced, when the sign function defining $Q$ is smeared to reflect experimental imprecision. For higher energy eigenstates, we find that LG violations diminish, showing the expected classicalization. With a $Q$ defined using a more general type of position coarse graining, we find two-time LG violations even in the ground state. We also show that two-time LG violations in a gaussian state are readily found if the dichotomic variable at one of the times is taken to be the parity operator. To demonstrate the versatility of the approach, we repeat much of the LG analysis for the Morse potential, finding qualitatively similar physical results. 

	\end{abstract}

	\maketitle

	\section{Introduction}
	
	Perhaps a great temptation within quantum mechanics is to ascribe the spookier extrapolations of its mathematics to phenomena which, although beyond our knowledge, are fundamentally classical in nature. This encourages us to  try to understand -- and place constraints on -- what the universe may be doing in between the snapshots we take of it in the form of measurements.  This endeavour, although foundational in nature, is becoming ever more relevant -- quantum technologies rely upon non-classical behaviour, and so tests which can verify the presence of truly quantum behaviour are very valuable. 
	
	One of the key open questions in this area is whether quantum coherence may persist up to the macroscopic level, which led to the introduction of the Leggett-Garg (LG) inequalities \cite{leggett1985, leggett1988,leggett2008,emary2013}.  These inequalities test a precise realist world-view known as macrorealism (MR), which is defined by the conjunction of three realist tenets -- that a system at all instants of time is definitely in one of the states available to it, that said state may be measured without influencing future dynamics of the system, and that future measurements do not affect past ones.  The violation of these inequalities imply a failure of MR, and thus, loosely speaking a detection of non-classical behaviour.
	
	The LG inequalities are typically formulated for a dichotomic observable $Q$, with possible values $s_i=\pm1$, which is measured at a sequence of times $t_1, t_2, \ldots, t_i$ . From these measurements, we can construct a data set consisting of the single time averages $\ev{Q_i}$, and the correlators $C_{ij}$, defined through
	\begin{equation}
	C_{ij} = \ev{Q_i Q_j} = \sum_{i,j}s_i s_j p(s_i, s_j),
	\label{eq1.1}
	\end{equation}
	where $Q_i$ denotes $Q(t_i)$.
	In keeping with the tenets of MR, the correlator must be measured in a non-invasive way, typically using ideal negative measurements \cite{knee2012} but other methods are also used \cite{halliwell2016a,majidy2019a,majidy2019b}.
	In Eq.(\ref{eq1.1}), $p(s_i, s_j)$ is the two-time probability describing the likelihood of obtaining results $s_i$ and $s_j$ and times $t_i$ and $t_j$.  We then consider making measurements at times $(t_1, t_2)$, $(t_2, t_3)$ and $(t_1, t_3)$, which yields three two-time probabilities.  For systems which obey the assumptions of MR, these three probabilities can always be matched as marginal probabilities of an underlying joint probability $p(s_1, s_2, s_3)$ and the correlators will obey the 
	three-time LG inequalities ($\LG{3}$):
	\bea
	L_1=1 + C_{12} + C_{23} + C_{13} & \ge & 0,
	\label{LG1}
	\\
	L_2=1 - C_{12} - C_{23} + C_{13} & \ge & 0,
	\label{LG2}
	\\
	L_3=1 + C_{12} - C_{23} - C_{13} & \ge & 0,
	\label{LG3}
	\\
	L_4=1 - C_{12} + C_{23} - C_{13} & \ge & 0.
	\label{LG4}
	\eea
	The same arguments can be made for measurements at just two times $t_1, t_2$,
	which leads to the two-time LG inequalities ($\LG{2}$)  \cite{halliwell2016b,halliwell2017,halliwell2019b,halliwell2019}:
	\beq
	1+s_1\ev{Q_1}+s_2 \ev{Q_2}+s_1 s_2 C_{12}\geq 0.
	\label{LG22}
	\eeq
	On their own, the $\LG{3}$ inequalities Eqs.~(\ref{LG1}-\ref{LG4}) form a set of necessary conditions for macrorealism -- although their violation indicates a failure of macrorealism, if they are satisfied, macrorealism does not necessary hold.  It is only by augmenting them with the set of twelve $\LG{2}$ inequalities, that we reach the complete set of necessary and sufficient conditions for macrorealism  \cite{halliwell2016b, halliwell2017,halliwell2019b, halliwell2019}.  The implications of this are two-fold.  Firstly, to test if a given data set admits a macrorealistic description, the entire set of $\LG{2}$ and $\LG{3}$ inequalities must be tested.  Secondly, a violation of just one $\LG{2}$ inequality is sufficient to show a failure of MR, which involving just two measurement times, may well be logistically simpler than testing the $\LG{3}$ inequalities.
	%
	
	
	
	Although originally proposed within the context of macroscopic quantum coherence, experimental LG tests have been nearly exclusively conducted on the discrete properties of microscopic systems; such as photons, silicon impurities, nuclear magnetic resonance, neutrino oscillations, and position superpositions in quantum random walks  \cite{majidy2019a,knee2012,goggin2011,formaggio2016, robens2015}.  For a thorough review of the LG inequalities, and experimental tests thereof, see Ref.~\cite{emary2013} and for a critique see  Ref.  \cite{maroney2014}.
	Some LG tests have come close to macroscopicity 
	\cite{knee2016}.
	There have also been some recent macroscopic non-classicality tests
	based on interferometric experiments, with quantum interference observed in $C_{60}$ and $C_{70}$ molecules \cite{arndt1999,arndt2001}, and in large organic molecules of masses up to 6910 amu \cite{gerlich2011}. These are not specifically LG tests, but proposals have been made to adapt them for this purpose \cite{pan2020,halliwell2021}.
	
	Most macroscopic systems that have been investigated with a view to exhibiting quantum coherence effects have the feature that they are described by continuous variables. Hence to approach LG tests in the macroscopic domain, it is natural to develop such tests for continuous variables, using, for example, a dichotomic variable $Q$ defined by simple coarse graining of position.
	In this work, we therefore pursue  an investigation into one of the most ubiquitous continuous variable systems -- the quantum harmonic oscillator (QHO), and bound systems more generally.
	
	
	Our theoretical investigation is in part inspired by the LG experiment proposed by Bose et al\cite{bose2018}. They used the dichotomic variable $Q={\rm sgn}(x)$, and focused on a single coherent state of the QHO.  By contrast with other tests for non-classicality, this approach does not require the fabrication of non-Gaussian states \cite{romero-isart2011}, nor the coupling to an ancillary quantum system \cite{scala2013,asadian2014}.   The work of Bose et al suggests that these quantum correlations persist, and remain experimentally feasible to detect, at scales of up to $10^6$-$10^9$ amu, meaning LG tests on the QHO are likely to push the limits of our observations of macroscopic coherence.  
	
	To further this programme, we derive analytical results for the temporal correlators for the QHO in a variety of experimentally accessible states, primarily energy eigenstates and superpositions thereof.
	We also derive an approximation for the temporal correlation functions which is applicable to any quantum system for which the energy eigenspectrum is (exactly or numerically) known.
	We determine substantial areas in the various parameter spaces in which combinations of LG2, LG3 and LG4 inequalities are satisfied or violated, thereby preparing the ground for experimental tests.
	
	In Section \ref{sec:calc}, we set up the problem and calculate temporal correlators working within the energy eigenbasis.  We develop a powerful technique for studying the partial overlap of energy eigenfunctions, which enables us to calculate the temporal correlators for any system with known energy eigenspectrum.  We apply this to the QHO, which yields a simple and useful approximation for the temporal correlator in the first excited state as a simple a cosine (just like the simple spin model used in LG tests).  We also demonstrate a procedure to calculate the exact temporal correlators for any energy eigenstate of the QHO.
	
	In Section \ref{sec:1and0}, we  explore the LG violations present in the QHO.  We find significant violations in the first excited state, and see that by creating superpositions of the ground state and the first excited state, we can find substantial regimes where the $\LG{2}$s and $\LG{3}$s are independently satisfied or violated. We find significant LG violations persist even with significant smoothing of projectors.

	In Section \ref{sec:higher}, we analyze the LG inequalities in the higher excited states of the QHO, and observe the expected classicalization, with the magnitude of LG violations rapidly decreasing as energy increases.

	In Section \ref{sec:mlvl}, we use a more general dichotomic observable, defined for arbitrary regions of space.  Using this more general variable we find LG violations even in the ground state.  
	
	In Section \ref{sec:morse}, to demonstrate the versatility of the techniques developed in Section \ref{sec:calc}, we calculate temporal correlators for the Morse potential, and perform a brief LG analysis, finding significant LG3 and LG4 violations.  We summarize and conclude in Section \ref{sec:conc}.
	
	The fine details of the complicated machinery required to calculate temporal correlators in the QHO, are largely relegated to a series of technical appendices.
	
	\section{Temporal correlators in bound systems}\label{sec:calc}
	
	\subsection{Conventions and Strategy}
	
	For most of this paper we work with the simplest choice of dichotomic variable $ Q=\text{sgn}(x)$, 
	however the method we develop also gives the result for more general observables involving arbitrary regions of space.  This is investigated further in Section ~\ref{sec:mlvl}.
	We will work with a general bound one-dimensional system with Hamiltonian $\hat H$ and energy eigenstates $\ket{n}$ with corresponding eigenvalues $E_n$.  
	
	We now outline our strategy.  We make use of the two-time quasi-probability, introduced in Ref.  \cite{halliwell2016b}, given by,
	\beq
	\label{eq:qp}
	q(s_1, s_2)=\text{Re} \Tr (P_{s_2}(t_2)P_{s_1}(t_1)\rho),
	\eeq
	where $s=\pm1$, and 
	\beq
	\label{eq:proj}
	P_s=\frac{1}{2}(1+ s \hat Q)=\theta(s \hat x),
	\eeq
	where $\theta(x)$ is the Heaviside step function. Time dependence is then handled in the Heisenberg picture.  
	
	We will use the quasi-probability in two ways. Firstly, it is proportional to the two-time LG inequalities, so negativity signifies an LG2 violation  \cite{halliwell2016b}. (The quasi-probability has a maximum negative value of $-\tfrac{1}{8}$).
	Secondly, it yields a simple way to extract the temporal correlators needed for the three-time LG inequalities, via its useful moment expansion
	\beq
	\label{momexp}
	q(s_1, s_2)=\frac{1}{4}\left(1+s_1 \expval{\hat Q_1}+s_2 \expval{\hat Q_2}+s_1 s_2 C_{12}\right).
	\eeq
	We will hence first calculate the quasi-probability, and then go on to extract the temporal correlators, which we can then use in the LG inequalities Eqs.~(\ref{LG1}-\ref{LG4}).  
	In contrast to the computation of correlators in simple spin models, these calculations are quite non-trivial.
	
	More detailed properties of the quasi-probability may be found in Ref.\cite{halliwell2016b}, but we make a few brief comments here. As indicated, it is used here purely as a mathematical tool and has no particular conceptual role beyond the fact that it is proportional to the two-time LG inequalities. It is clearly non-negative for commuting pairs of projections, so any negativity is associated with incompatibility of the pairs of measurements. It can usefully be expressed in terms of the Wigner-Weyl representation\cite{halliwell2019c}, from which one can see that it can be negative even when the Wigner function of the initial state is non-negative, as is the case for gaussian initial states. We shall see an example of this later in this paper, and see also  Refs.\cite{halliwell2019c, halliwell2021}. The quasi-probability is also very closely related to the two-time sequential measurement formula
	\begin{equation}
	p(s_1,s_2) = \Re \Tr \left(  P_{s_2} (t_2) P_{s_1} (t_1) \rho P_{s_1} (t_1)   \right),   
	\end{equation}
	This is always non-negative and has a moment expansion similar in form to Eq.(2.3), with the only difference that $ \langle \hat Q_2 \rangle$  is replaced with $ \langle \hat Q_2^{(1)} \rangle$,  
	the average of $Q_2$ in the presence of an earlier measurement at $t_1$ which has been summed out. (See Ref.\cite{halliwell2016b}). Note in particular that both the quasi-probability and sequential two-time measurement formula have the same correlator, which has the quantum-mechanical form
	\begin{equation}
	\label{eq:qcorr}
	C_{12} =  \frac{1}{2} \langle \hat Q_1 \hat Q_2 + \hat Q_2 \hat Q_1 \rangle.
	\end{equation}
	Physically, this corresponds to the fact that the same correlator is obtained using a pair of measurements in which the first measurement is projective or weak \cite{halliwell2016b, fritz2010}, or more generally, any one of a family of ambiguous measurements intermediate between these two options (as can be seen in Ref.\cite{halliwell2019c}). The relevant non-invasive measurement protocols for LG tests of these quantities are described in more detail in Refs.\cite{halliwell2016b, halliwell2017,halliwell2019b, halliwell2019a}.

	
	\subsection{Calculating the quasi-probability}
	Using the definition of the two-time quasi-probability Eq.(\ref{eq:qp}), and Eq.(\ref{eq:proj}), we have
	\beq
	\label{qp2}
	q(s_1, s_2)=\text{Re} \Tr (e^{\frac{iHt_2}{\hbar}}\theta(s_1 \hat x )e^{-\frac{i H(t_2-t_1)}{\hbar}}\theta (s_2\hat x )e^{\frac{-iHt_1}{\hbar}}\rho).
	\eeq
	We will often work with the difference between measurement times, and so introduce the variable $\tau = t_2-t_1$.
	
	Initially we calculate one element of the quasi-probability $q(+,+)$, where the full quasi-probability $q(s_1, s_2)$ may often be reconstructed through symmetry arguments.  Working within the energy eigenbasis, we write Eq.(\ref{eq:qp}) as
	\beq
	q(+,+)=\Re\sum_{m=0,n=0}^{\infty}\!\!\!\braket{\psi}{m}\!\!\!\mel{m}{P_+(t_2)P_{+}(t_1)}{n}\!\!\braket{n}{\psi},
	\eeq
	which equals
	\beq
	q(+,+)=\Re \sum_{m=0,n=0}^{\infty}\braket{\psi}{m}\braket{n}{\psi}q_{mn},
	\eeq
	where
	\beq
	\label{eqn:qmn2}
	q_{mn}=
	\mel**{m}{e^{\frac{i H t_2}{\hbar}}\theta(\hat x)e^{-\frac{iH\tau}{\hbar}}\theta(\hat x)e^{-\frac{i H t_1}{\hbar}}}{n}.
	\eeq
	Using the position representation for the step functions $\theta(\hat x)=\int^\infty_0 \vert x \rangle \langle x \vert \mathop{dx}$, this may be written explicitly as
	\beq
	\label{eq:qmn}
	q_{mn}=e^{i\frac{E_m}{\hbar}t_2-i\frac{E_n}{\hbar}t_1}\int_{0}^{\infty}\int_{0}^{\infty}\mathop{dx}\mathop{dy}\braket{m}{x}\!\!\mel{x}{e^{-\frac{i H \tau}{\hbar}}}{y}\!\braket{y}{n},
	\eeq
	where from here the calculation forks, with two obvious ways to proceed. One approach is to insert a resolution of unity in the energy eigenbasis, and truncate the infinite sum to yield an approximation to the correlators, which we will proceed with first.  The second approach is to insert the expression for the propagator, if known, which will allow us to calculate exact expressions for the quasi-probability.
	
	The single time averages are given by 
	\beq
	\ev*{\hat{Q}}=\ev{\sgn(\hat x)}{n}.
	\eeq
	In the special case of symmetric potentials, since $\text{sgn}(\hat x)$ flips the parity of $\ket n$, this represents the overlap between an odd and an even state. Hence in this case, the single time averages in the moment expansion Eq.(\ref{momexp}) are identically zero.  This yields a simple relationship between the correlator and the quasiprobability,
	\beq
	\label{eq:qeig}
	q(+,+)=\frac{1}{4}(1+C_{12}).
	\eeq
	We adopt the notation $C_{12}^{\ket{n}}$, to denote the temporal correlator between times $t_1$ and $t_2$, for a given energy eigenstate $\ket{n}$.
	
	\subsection{Quasi-probability for energy eigenstates}
	\label{subsec:inf}
	We use the approximation approach first, whereupon inserting the resolution of unity, we have
	\beq
	q_{mn}=e^{i\frac{E_m}{\hbar}t_2-i\frac{E_n}{\hbar}t_1}\sum_{k=0}^{\infty}e^{-i \frac{E_k}{\hbar}(t_2-t_1)}\int_{0}^{\infty}\int_{0}^{\infty}\mathop{dx}\mathop{dy}\braket{m}{x}\!\braket{x}{k}\!\braket{k}{y}\!\braket{y}{n}.
	\eeq
	The integration is now separable into two integrals, the partial overlap of energy eigenstates,
	\begin{equation}
	J_{k\ell}=\int_0^{\infty}\mathop{dx}\braket{k}{x}\braket{x}{\ell}.
	\end{equation}
	Surprisingly, as detailed in Appendix \ref{app:wronski}, these integrals can be completed for generic potentials over arbitrary boundaries $x_1$, $x_2$, with result
	\begin{equation}
	\label{eqn:wronski}
	J_{k\ell}(x_1, x_2)=\frac{1}{2(\varepsilon_\ell- \varepsilon_k)}\left[\psi_k'(x_2)\psi_\ell(x_2)-\psi_\ell'(x_2)\psi_k(x_2)-\psi_k'(x_1)\psi_\ell(x_1)+\psi_\ell'(x_1)\psi_k(x_1)\right],
	\end{equation}
	where $\psi_k(x)=\braket{x}{k}$.  Hence the matrix elements of the quasi-probability are
	\beq
	q_{mn}=e^{i\frac{E_m}{\hbar}t_2-i\frac{E_n}{\hbar}t_1}\sum_{k=0}^{\infty}e^{-i\frac{E_k}{\hbar}(t_2-t_1)}J_{mk}J_{nk}.
	\eeq
	The quasi-probability for a single energy eigenstate is the real part of the diagonal elements,
	\beq
	\label{eqn:qpapprox}
	q_n(+,+)=\Re e^{i \frac{E_n}{\hbar} \tau }\sum_{k=0}^{\infty}e^{-i \frac{E_k}{\hbar}\tau }J_{nk}^2.
	\eeq

	By truncating this sum, we are able to calculate the quasi-probability for any coarse-graining of space, and for any soluble bound system, purely in terms of the spectrum of its Hamiltonian.  We will primarily use this result with the QHO in this work, however in Sec.~\ref{sec:morse}, we consider its application to other systems.  Further details about implementation of this formula, including a truncation error estimate $\Delta_n(m)$ defined in Eq.(\ref{eq:trunc}), are found in Appendix \ref{app:calcdeets}.
	
	We also note in passing that Eq.(\ref{eqn:wronski}) unlocks the result of several indefinite integrals in the form of partial spatial overlaps of special functions, which we have not been able to find in the literature. These include; generalised Laguerre polynomials (Morse potential \cite{dahl1988}), Mathieu functions (quantum pendulum \cite{galvez2019}), Airy functions (triangular potential \cite{khonina2013}), and Hypergeometric functions (transformation technique \cite{morales2015}), with the exactly soluble potential leading to the result bracketed.

	\subsection{Application to the QHO}
	
	The results so far have been general to any bound potential.  We now apply them to the systems defined exactly (or approximately) by the harmonic oscillator Hamiltonian,
	\beq
	\hat H=\frac{\hat p_{\rm{phys}}^2}{2m}+\frac12 m\omega^2 \hat x_{\rm{phys}}^2,
	\eeq
	with physical position and momentum $x_{\rm{phys}}$ and $p_{\rm{phys}}$.  We continue to denote energy eigenstates by $\ket n$, and use natural units, so we work with dimensionless position and momentum $x$ and $p$, defined by $\sqrt{\hbar/(m\omega)} x=x_{\rm{phys}}$ and $p \sqrt{(m\omega)/\hbar}=p_{\rm{phys}}$.  We will similarly write energy in terms of $\hbar \omega$, with $E_n=(n+\frac12)\hbar \omega=\varepsilon_n \hbar \omega$.  The energy eigenstates within the position basis are then given by
	\beq
	\braket{n}{x}=\frac{1}{\sqrt{2^n n!}}\pi^{-\frac{1}{4}}\exp(-\frac{1}{2}x^2)H_n(x),
	\eeq
	where $H_n(x)$ are the Hermite polynomials.
	
	For the ground state, the infinite series needs many terms to reach good convergence, and so we instead turn to $q_{11}$, whereupon taking the real part, we have the quasi-probability $q(+,+)$, for the first excited state.  We calculate the first few $J_{nm}$ using Eq.(\ref{eq:jijlim}). We find
	$J_{1,0}^2=\frac{1}{2\pi}$, $J_{1,1}^2=\frac14$, $J_{1,2}^2=\frac{1}{4\pi}$, $J_{1,3}^2=0$, and $J_{1,4}^2=\frac{1}{48\pi}$.  Using just the first three terms leads to a very good approximation to the correlator, with $\Delta_1(2)=0.011$.  The sum may be explicitly written as  $	q_{11}=\frac{1}{4}+\frac{1}{4\pi}e^{-i\omega\tau}+\frac{1}{2\pi}e^{i\omega \tau}$. Taking the real part, we find
	\beq
	\label{eq:approx}
	q(+,+)=\frac{1}{4}+\frac{3}{4\pi}\cos \omega \tau.
	\eeq
	By comparing Eq.(\ref{eq:approx}) to the expression for the quasi-probability of a pure eigenstate Eq.(\ref{eq:qeig}), we can identify the correlator as 
	\beq
	\label{eq:cos}
	C_{ij}^{\ket1}\approx\cos\omega\tau,
	\eeq
	where we have dropped the coefficient of $\tfrac{3}{\pi}\approx0.955$, by the requirement that $C_{ij}\to 1$ as $\tau \to 0$.
	
	This approximation is interesting since it says that for the first excited state, the QHO to a good approximation behaves just like the canonical simple spin-$\frac{1}{2}$ example used in much LG research.  We hence may borrow intuition from this simpler system, and we know to expect LG violations.  
	This similarity can be understood by explicitly calculating the temporal correlator using the three-term approximation to the post-measurement state.  The temporal correlator here is defined as
	\begin{equation}
	C_{12}^{\ket1}=\Re \expval{\hat Q_2 \hat Q_1}{1}.
	\end{equation}
	From Eq.(\ref{eq:proj}), we have $\hat Q=2\theta(\hat x)-1$, so using the three-term approximation and taking $t_1=0$, we find
	\begin{equation}
	\hat Q_1 \ket{1}\approx\frac{2}{\sqrt{2\pi}}\ket0+\frac{1}{\sqrt{\pi}}\ket2.
	\end{equation}  
	Similarly, taking $t_2=\tau$, we find
	\begin{equation}
	\bra{1}\hat Q_2 \approx e^{i\omega\tau}\left(\frac{2}{\sqrt{2\pi}}\bra0+e^{-2i \omega\tau}\frac{1}{\sqrt{\pi}}\bra2.\right).
	\end{equation}
	Hence like the spin-$\frac{1}{2}$ case, it boils down to just two states.
	Contracting these result, and taking the real part yields
	\begin{equation}
	C_{12}^{\ket1}=\frac3\pi \cos(\omega \tau) \approx \cos(\omega \tau) 
	\end{equation}
	
	We note that taking the next order of approximation we have $\Delta_1(4)=0.005$, and clearly a much better approximation.  The expression for the correlator (again normalized to 1 as $\tau\to0$), is then
	\beq
	C_{ij}^{\ket1}\approx \frac{36}{37}\cos \omega \tau + \frac{1}{37}\cos 3\omega \tau.
	\eeq
	
	\subsection{Exact Correlators for the QHO}
	For the QHO, we are also able to calculate exact expressions for the correlators,  whereupon inserting the expression for the QHO propagator into Eq.(\ref{eq:qmn}), we find
	\begin{multline}
	\label{eq:qmnfull}
	q_{mn}=\mathcal{N}_{mn}(\tau) e^{i\frac{E_m}{\hbar}t_2-i\frac{E_n}{\hbar}t_1}\times\\\int^{\infty}_0\int^{\infty}_0 \mathop{dr}\mathop{ds}H_m(r)H_n(s)e^{-\frac{r^2}{2}}e^{-\frac{s^2}{2}}\exp({i \frac{1}{2\tan(\omega \tau)}(r^2+s^2)- i\frac{1}{\sin(\omega \tau)}rs}),
	\end{multline}
	where $\mathcal{N}_{mn}(\tau)$ is a dimensionless normalisation factor, defined by
	\beq
	\mathcal{N}_{mn}(\tau)=\frac{1}{\pi}\frac{1}{\sqrt{2^n n!}}\frac{1}{\sqrt{2^m m!}}\frac{1}{\sqrt{2i \sin(\omega\tau)}}
	\eeq
	This integral can be completed, via a generating integral approach, which is detailed in Appendix~\ref{app:gen}.  Using this result, we are able to write the exact results for the temporal correlators.  To do so, it is helpful to introduce the function
	\beq
	f(\tau)=-ie^{-i\frac{\omega\tau}{2}}\sqrt{2i\sin\omega\tau}.
	\eeq
	The correlators for the groundstate and first excited state may then be written as
	\begin{align}
	\label{eq:arctan}
	C_{ij}^{\ket0}&=\frac{2}{\pi}\Re\arctan\left(\frac{1}{f(\tau)}\right),\\
	\label{cor:1s}
	C_{ij}^{\ket1}&=\frac{2}{\pi}\Re\left[\arctan\left(\frac{1}{f(\tau)}\right)+f(\tau)\right].
	\end{align}
	These two rather non-intuitive expressions are plotted in Fig. \ref{fig:corrs}, where they are easily seen to have the broadly expected physical behaviours
	The expressions for the first nine correlators are included in Appendix~\ref{app:corrs}. 
	\subsection{Matrix Elements for a two-state oscillator}
	It will turn out that very much can be studied in the QHO, with initial superposition states of just the ground state and first excited state.  That is working with states of the form 
	\beq
	\label{eq:state}
	\ket \psi = a \ket 0 + b \ket 1.
	\eeq
	To determine the off-diagonal elements of $q_{mn}$ defined in Eq.(\ref{eqn:qmn2}), it is simplest to do so term-by-term in the moment expansion Eq.(\ref{momexp}).
	
	The quantum mechanical temporal correlator is given by Eq.(\ref{eq:qcorr}), with matrix elements
	\beq
	\mel{m}{\hat C_{ij}}{n}=\frac12\mel*{m}{\hat Q_1 \hat Q_2 + \hat Q_2 \hat Q_1}{n}.
	\eeq
	We now note that  $\hat Q_1 \hat Q_2 + \hat Q_2 \hat Q_1$ is invariant under reflections. This means that $(\hat Q_1 \hat Q_2 + \hat Q_2 \hat Q_1)\ket{n}$ must have the same parity as $\ket{n}$.  Hence in cases where $m$ is odd and $n$ even, or vice-versa, the correlator represents the overlap of an even state with an odd state, and is therefore zero in these cases.  Hence for the general state Eq.(\ref{eq:state}), the correlator is simply given as the mixture
	\beq
	\label{supcor}
	C_{ij}=\abs{a}^2 C_{ij}^{\ket0}+ \abs{b}^2 C_{ij}^{\ket1}.
	\eeq
	We now determine the matrix elements of the quasi-probability.  Owing to the argument preceding Eq.(\ref{eq:qeig}), the diagonal elements take the simple form 
	\beq
	q_{nn}(s_1, s_2)=\frac{1}{4}\left(1+s_1 s_2 C_{ij}^{\ket n}\right).
	\eeq
	Similarly, with the vanishing of the correlators on the off-diagonals, we see that
	\beq
	q_{01}(s_1, s_2)=\frac{1}{4}\mel{0}{s_1 \hat Q_1+s_2 \hat Q_2}{1}.
	\eeq
	As these averages involve only a single time each, the time dependence is trivial, and we just require the value of
	\beq
	\ev*{\hat{Q}}=\mel{0}{\text{sgn}(\hat x)}{1},
	\eeq
	which is readily calculated using Eq.(\ref{eq:jijlim}) as $\ev*{\hat{Q}}=\sqrt{\frac{2}{\pi}}$.

	We hence find the two-time quasi-probability for the state Eq.(\ref{eq:state}) to be
	\beq
	\label{eq:q2so}
	q(s_1, s_2)=\frac{1}{4}\left[1+s_1\left(2\sqrt{\frac{2}{\pi}}\Re a^*b e^{i\omega t_1}\right)+s_2\left(2\sqrt{\frac{2}{\pi}}\Re a^*b e^{i\omega t_2}\right)+s_1 s_2\left(\abs{a}^2C_{ij}^{\ket0}+\abs{b}^2C_{ij}^{\ket1}\right)\right].
	\eeq
	\subsection{The classical analogue}
	\begin{figure}
		\subfloat[]{{\includegraphics[height=5.1cm]{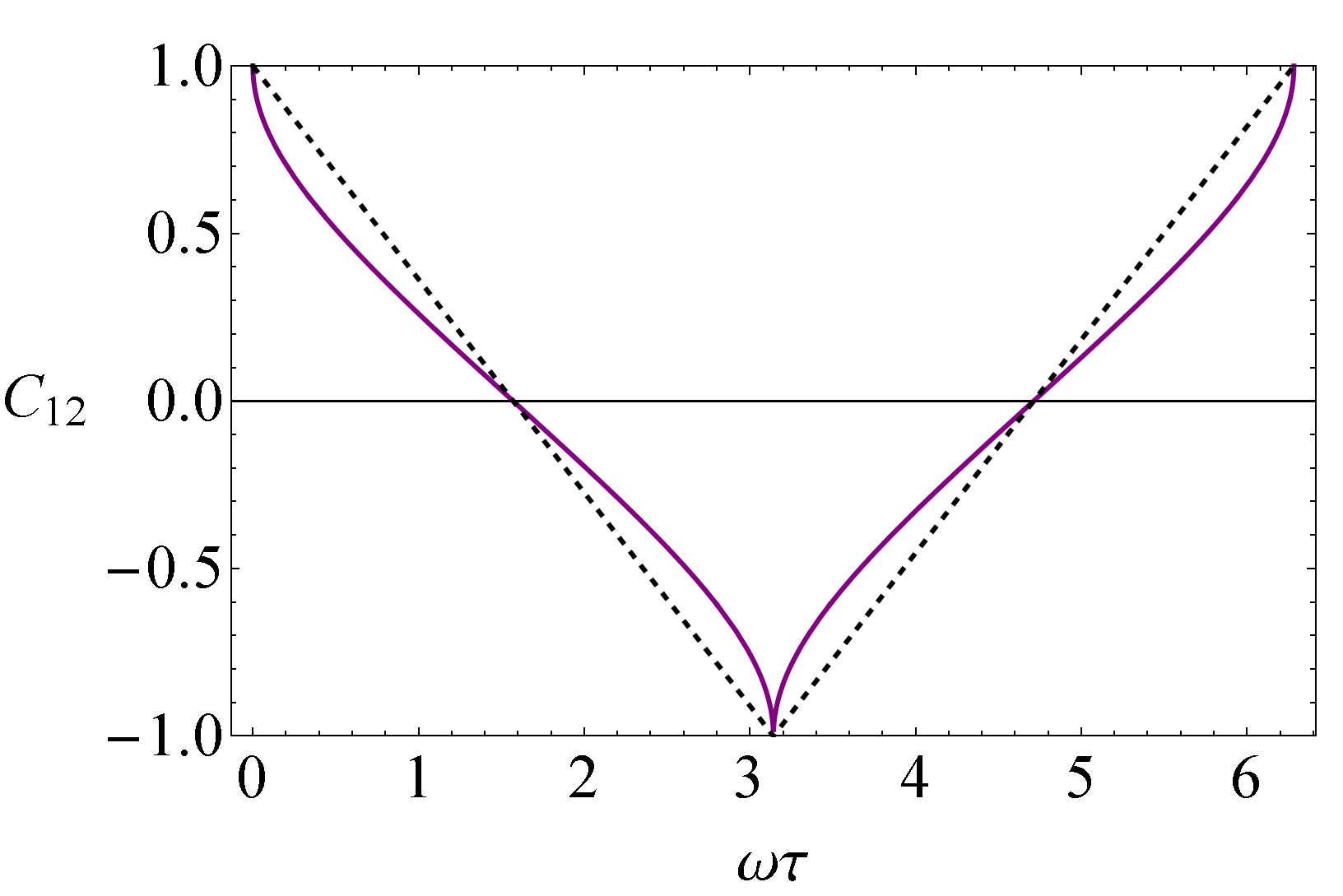}}}%
		\qquad
		\subfloat[]{{\includegraphics[height=5.1cm]{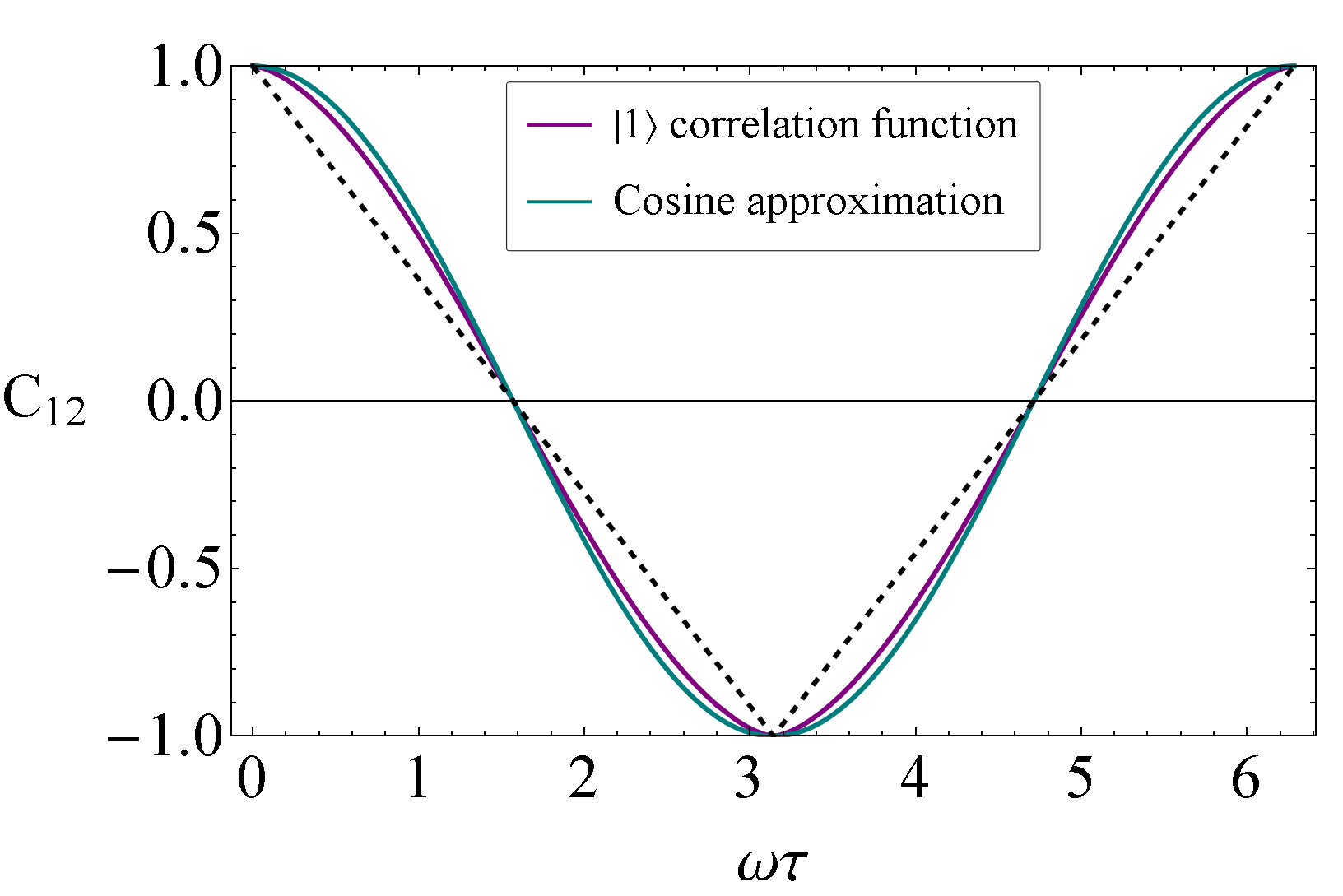}}}
		\caption{In (a), the temporal correlator for the ground state of the QHO Eq.~(\ref{eq:arctan}) is plotted, alongside the classical analogue Eq.~(\ref{eq:class}) (dashed).  In (b), the temporal correlator for the first excited state Eq.~(\ref{cor:1s}) is plotted, alongside the correlator for the simple spin case, showing the close similarity mentioned in the text.  The classical analogue is again shown (dashed).}%
		\label{fig:corrs}%
	\end{figure}
	It is useful in many of these calculations to compare the correlators obtained with their classical analogue. 
	This is readily found from the classical analogue of the quasi-probability, namely $\langle \theta (x) \theta (x(\tau) \rangle $, with phase-space initial state $f(x^2 + p^2)$ which is normalised to $2\pi\int_0^{\infty}\mathop{dr}r f(r)=1$.  This choice covers fixed energy states, and mixtures thereof.  We then have,
	\beq
	\mathbbm{q}(0, \tau)=\int_{-\infty}^\infty\int_{-\infty}^\infty \mathop{dx}\mathop{dp}f(x^2 + p^2)\theta(x) \theta(x \cos \omega \tau+p\sin \omega \tau).
	\eeq
	The step-functions are easiest handled in polar coordinates as
	\beq
	\mathbbm{q}(0, \tau)=\int_{0}^{\abs{\pi-\omega \tau}}\mathop{d\theta}\int_0^\infty \mathop{dr}r f(r),
	\eeq
	where inserting the normalisation condition on $f(r)$ yields
	\beq
	\label{eq:class}
	\mathbbm{q}(0, \tau)=\frac{\lvert{\pi-\omega \tau}\rvert}{2\pi},
	\eeq
	
	which has the corresponding classical correlator of $\mathbbm{C}_{12}=-1+\frac{2}{\pi}\lvert{\pi-\omega \tau}\rvert$.
	
	We now plot two of the QHO correlators, alongside the classical analogue, in Fig.~\ref{fig:corrs}.  We also plot the correlator for the first excited state, alongside the approximation Eq.(\ref{eq:cos}), which also serves to compare the behaviour of the first excited state of the QHO with the canonical simple spin model typically used in LG research. 
	
	\section{Leggett-Garg Violations in the two-state QHO}\label{sec:1and0}
	\subsection{Four Regimes}
	In this section, we study the LG inequalities in the scenario where we have access only to the ground state and the first excited state, the two-state oscillator.  
	This allows us to take full advantage of the simplicity of the expression for the correlator in this two-state superposition, Eq.(\ref{supcor}).  We will find that even in this state space, we have access to four regimes of interest, which are laid out in Table \ref{tbl:reg}.
	
	\begin{table}[!htbp]
		\def\arraystretch{1.125}
		{\setlength{\tabcolsep}{0.5em}
			\begin{tabular}{ Sc | Sc | Sc  }
				\textbf{Regime} & \textbf{LG3s} & \textbf{LG2s} \\
				\hline
				I & \includegraphics[scale=0.05]{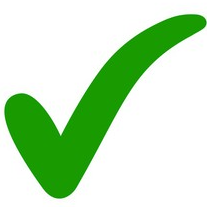} & \includegraphics[scale=0.05]{check} \\
				II & \includegraphics[scale=0.05]{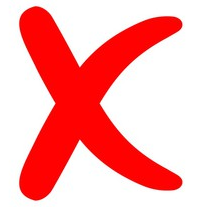} & \includegraphics[scale=0.05]{check}  \\
				III & \includegraphics[scale=0.05]{check} & \includegraphics[scale=0.05]{cross} \\
				IV & \includegraphics[scale=0.05]{cross} & \includegraphics[scale=0.05]{cross}
		\end{tabular}}
		\caption{The four regimes testable with the $\LG{2}$ and $\LG{3}$ inequalities.  A tick denotes that the complete set of inequalities is satisfied, whereas a cross indicates that one or more of the inequalities in that set is violated.}
		\label{tbl:reg}
	\end{table}
	
	These four regimes take their importance from the fact that, for the data set consisting of the three $\langle Q_i \rangle$ and three correlators,
	the LG inequalities form necessary and sufficient conditions for MR only when all the $\LG{2}$ and $\LG{3}$ inequalities are satisfied.  This has a couple of interesting consequences in cases where the $\LG{3}$ inequalities are satisfied.  From the sufficiency aspect; although typically just $\LG{3}$s are experimentally tested, their being satisfied does not in fact indicate the system behaves macrorealistically, since the $\LG{2}$s must also be verified as satisfied.  This means using solely the $\LG{3}$s to test for MR leaves space open for false positives.  From the necessity aspect; since all conditions must be satisfied, it means the violation of one or more of the $\LG{2}$s is enough to indicate the failure of MR, which since it involves just two times, may be easier to implement experimentally.
	These four regimes have previously been explored both experimentally and theoretically in simple spin models \cite{majidy2021a}.
	
	The QHO limited to the superpositions of the $\ket0$ and $\ket1$ states makes a complete playground for the experimentalist, exhibiting all variants of MR tested by the LG inequalities.  We analyze and provide examples of each of the regimes in Table~\ref{tbl:reg}. In what follows, we survey the parameter-space of these two-state superpositions to determine where each of the four regimes lie.
	\subsection{LG inequalities for pure eigenstates}

	We first consider initial states which are energy eigenstates. The LG2s are trivially satisfied for these states since $\ev{Q_i}=0$, so we can only access regimes I and II in Table \ref{tbl:reg}. This is not the case for superpositions.
	
	Consider now the $\LG{3}$s, $L_{i}^{\ket n}$, where $i$ indexes one of the four $\LG{3}$ kernels Eq.(\ref{LG1}-\ref{LG4}), with the initial state $\ket n $.  For example,
	\begin{equation}
	L_{1}^{\ket 0}=1+C_{12}^{\ket0}+C_{13}^{\ket0}+C_{23}^{\ket0}.
	\end{equation}
	
\begin{figure}[t]
		\subfloat[]{{\includegraphics[height=5.1cm]{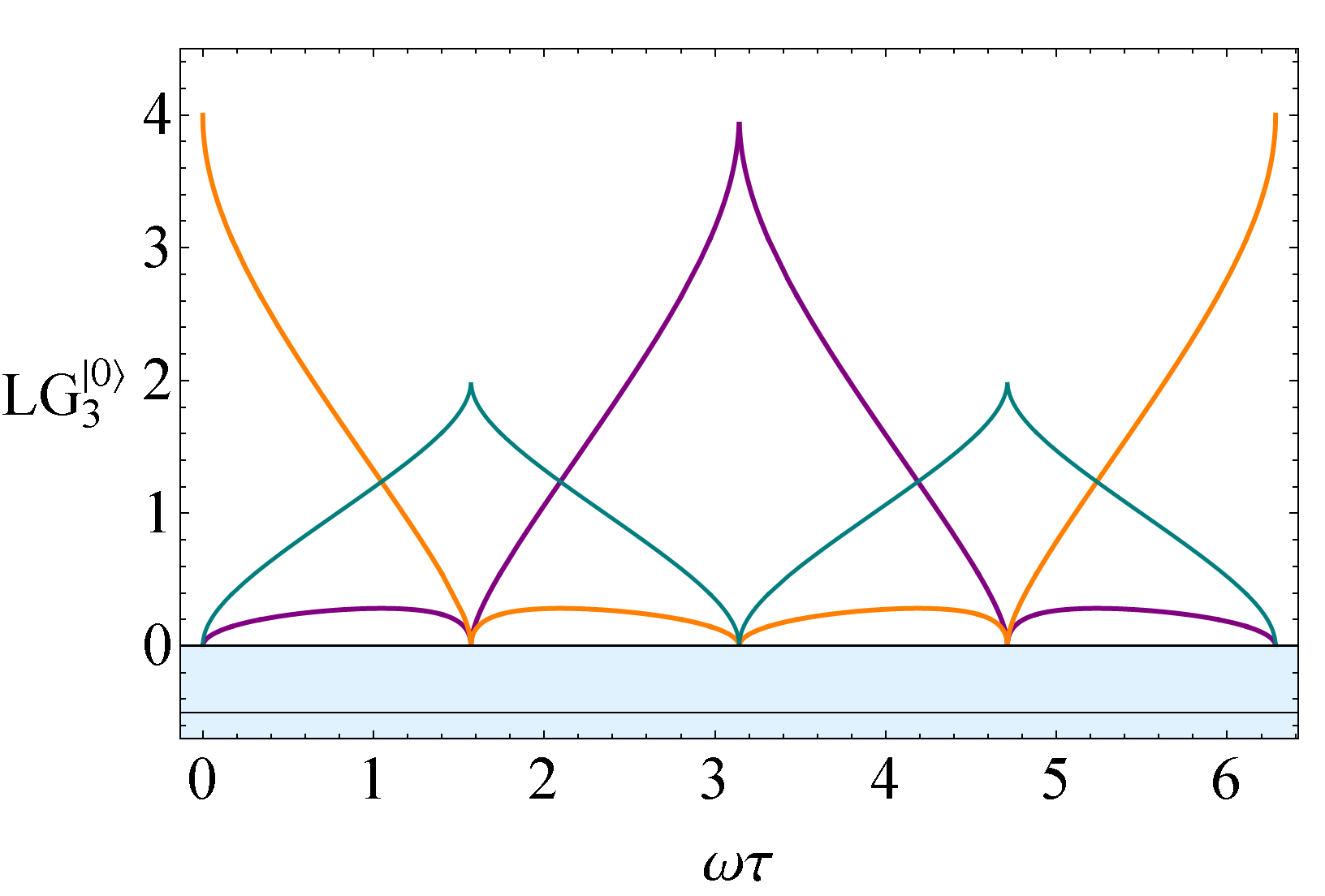}}}%
		\qquad
		\subfloat[]{{\includegraphics[height=5.1cm]{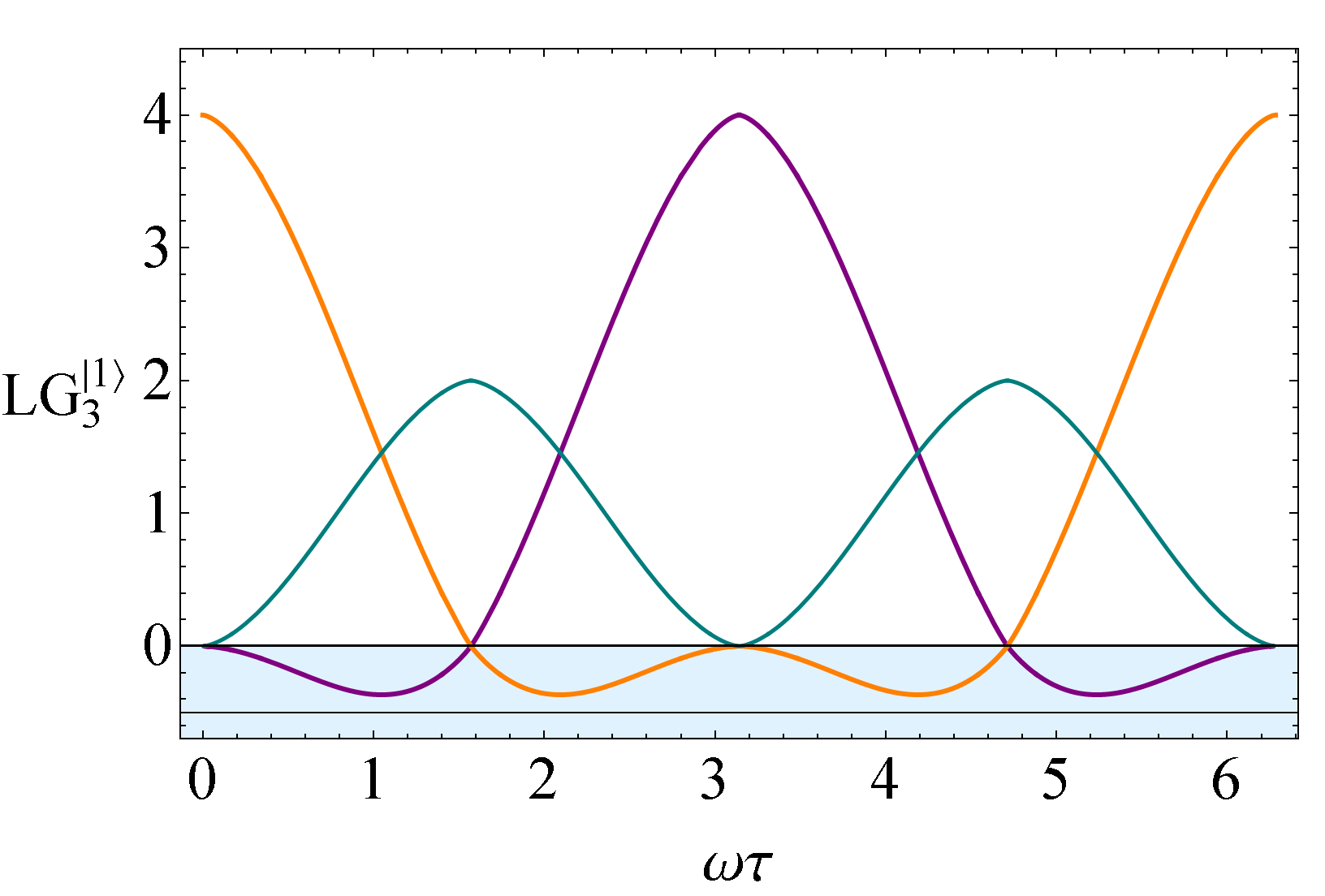}}}
		\caption{In (a), the three-time LG inequalities are plotted for the ground state of the QHO, where they are satisfied at all times.  In (b), the same inequalities are shown for the first excited state, where they are violated everywhere, except for points of measure zero.  These violations come close to the maximal violation, the L\"{u}ders bound of $-\frac12$.}%
		\label{fig:lgpure}%
	\end{figure}

	These four inequalities are plotted in Fig.~\ref{fig:lgpure}, (although two of them coincide).  For the ground state, the $\LG{3}$ inequalities are satisfied always.  Since the $\LG{2}$ inequalities are trivially satisfied as well, this means that for the ground state lies in regime I of Table \ref{tbl:reg}.  For the first excited state, we see the $\LG{3}$ are violated always (except at points of measure zero). The magnitude of the violation is significant, reaching approximately 73\% of the L\"{u}ders bound of $-\frac12$.  This means that the first excited state corresponds to regime II.
		
	\subsection{LG inequalities for superposition states}
	
	Since the matrix elements of the correlator are diagonal for
	superpositions of $\ket0$ and $\ket1$, the LG3s will simply be a convex sum of $L_i^{\ket0}$ and $L_i^{\ket1}$.  We can hence continuously transition between states with no $\LG{3}$ violation, and those with $\LG{3}$ violation everywhere.
	We begin by parametrising the two-state superpositions as
	\beq
	\label{eq:param}
	\ket{\psi}=\cos \tfrac{\theta}{2} \ket 0+e^{i\phi}\sin \tfrac{\theta}{2} \ket 1,
	\eeq
	where $0\leq \theta \leq \pi$, and $0\leq \phi \leq 2\pi$.

\begin{figure}
		{\includegraphics[height=6.4cm]{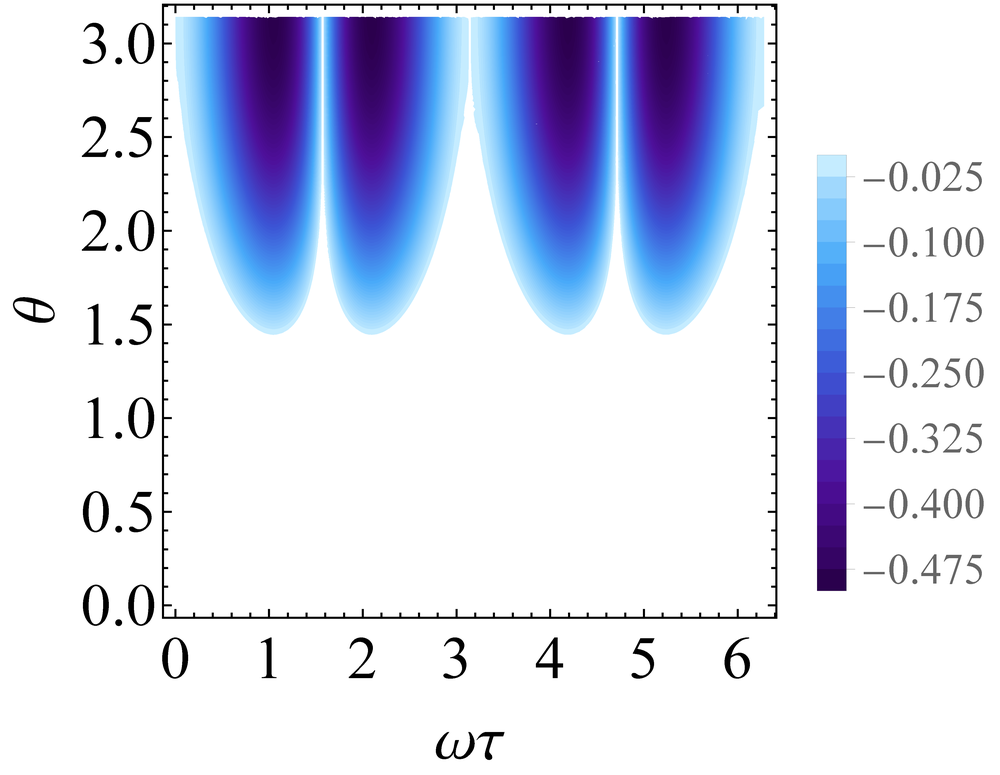}}
		\caption{The regions where at least one of the $\LG{3}$ inequalities is violated are shown as a function of time between measurements, and superposition coefficients.  The shading corresponds to the magnitude of the $\LG{3}$ violation.}%
		\label{fig:lgsup}%
	\end{figure}

	Since the $\LG{3}$ inequalities are constructed purely from correlators, using Eq.(\ref{supcor}), we find
	\beq
	L_i(\theta)=\cos^2\tfrac{\theta}{2} ~L_i^{\ket0}+\sin^2\tfrac{\theta}{2}~L_i^{\ket1},
	\eeq
and so a two-dimensional space of $\theta$ and $\tau$ defines whether the $\LG{3}$ inequalities for the superposition are satisfied.  This space is plotted in Fig.~\ref{fig:lgsup}, where for a given $\theta$, there are three possible cases; the $\LG{3}$ are violated for all $\tau$, the $\LG{3}$ are satisfied for all $\tau$, or a non-trivial mixture of the two, where they are satisfied for some ranges of $\tau$, and violated for others.

\begin{figure}
		\subfloat[]{{\includegraphics[height=7cm]{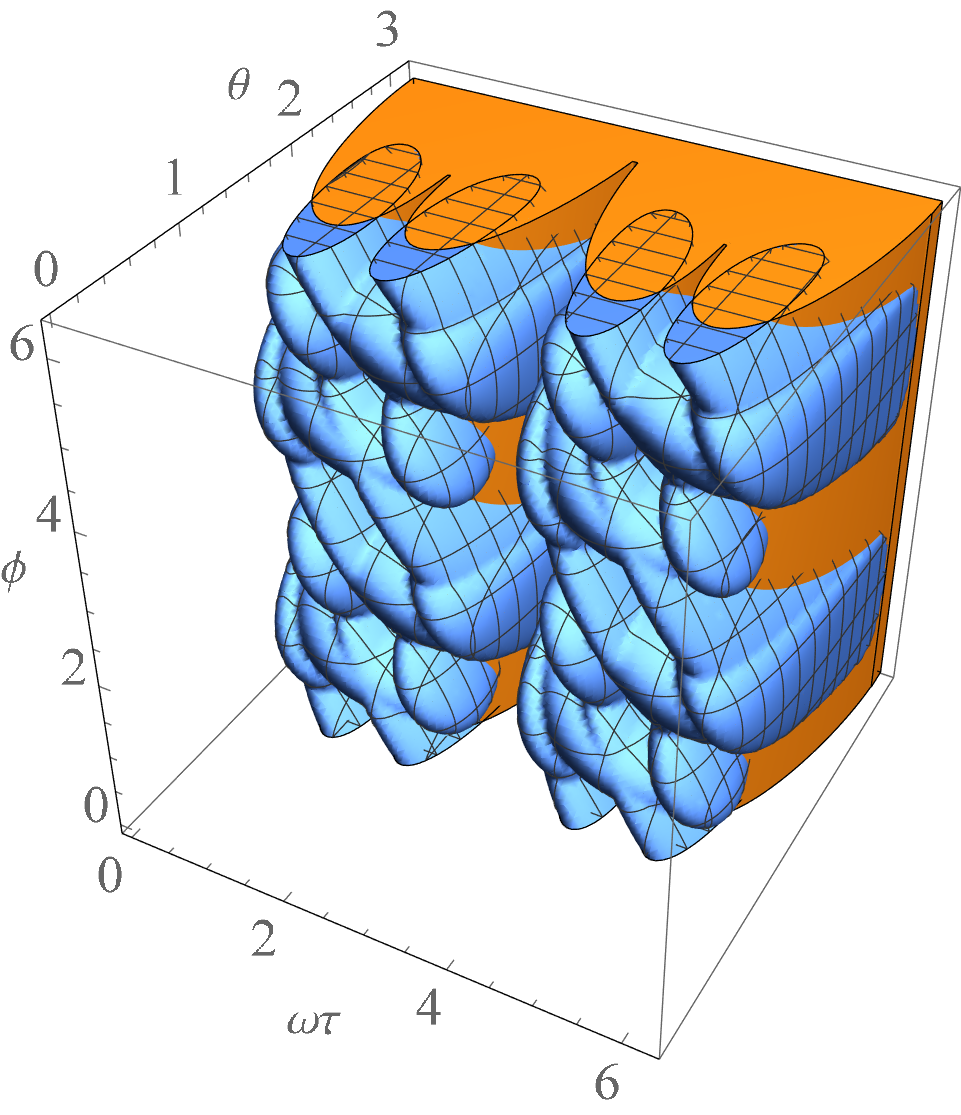}}}	%
		\qquad
		\subfloat[]{{\includegraphics[height=6.3cm]{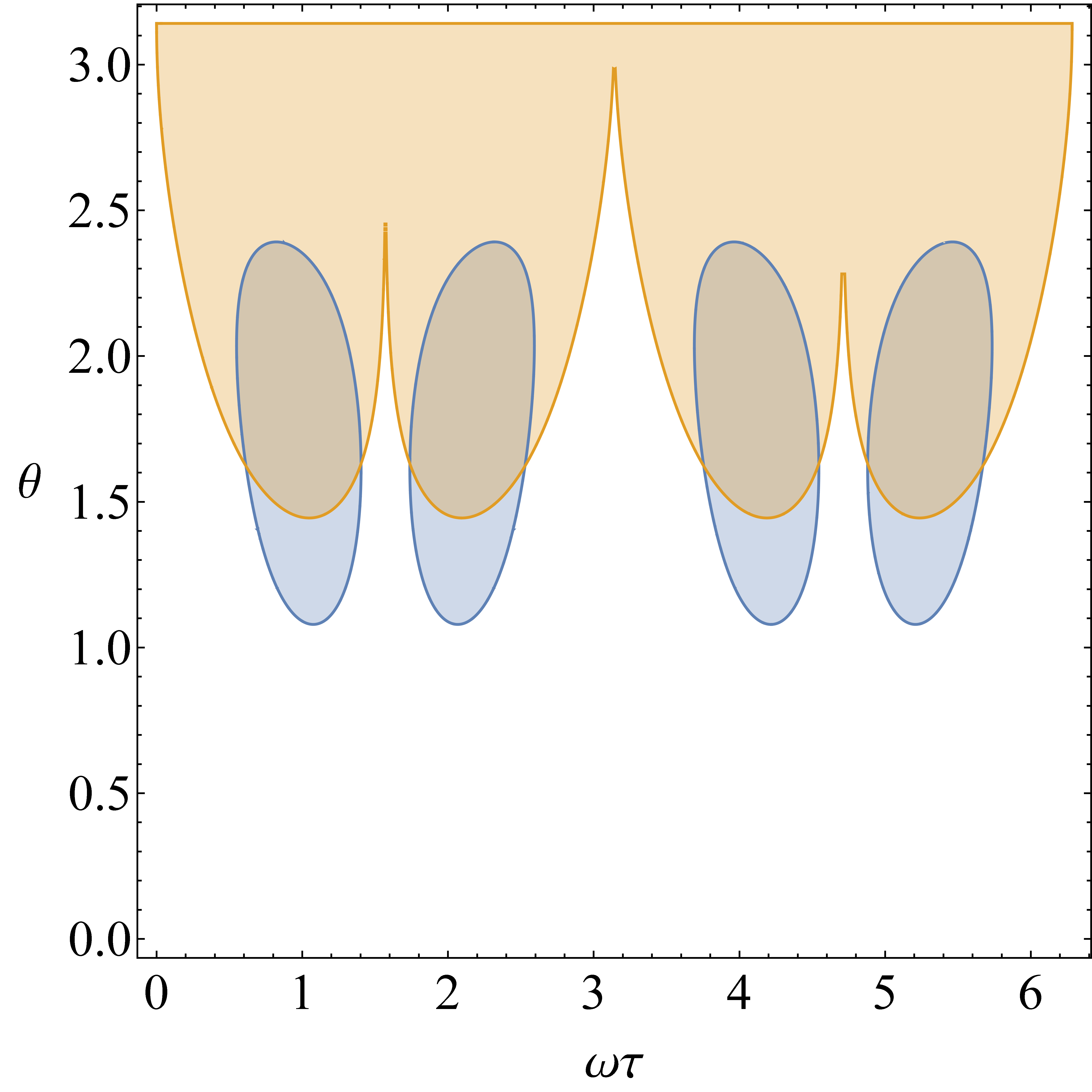}}}
		\caption{In (a) the behaviour of the $\LG{2}$ and $\LG{3}$ inequalities is plotted, for all possible superpositions of $\ket0$ and $\ket1$.  Within the orange (non-meshed) region, at least one of the four $\LG{3}$ inequalities are violated, and within the blue (meshed) region, at least one of the twelve $\LG{2}$ inequalities are violated.  In (b), a slice of this parameter space at $\phi=\pi$ is presented. }%
		\label{fig:lg2frac}%
	\end{figure}

	We now analyze the behaviour of the two-time LG inequalities.  Due to interference terms, the averages $\langle Q_i\rangle$ are non-zero for superposition states, which can produce $\LG{2}$ violations.  Using the parametrisation Eq.(\ref{eq:param}) in the expression for the quasi-probability Eq.(\ref{eq:q2so}), we find
	\begin{multline}
	q(s_1, s_2)=\frac14\Bigg[1+s_1\left(\sqrt{\frac{2}{\pi}}\sin \theta\cos (\phi+\omega t_1)\right)+s_2\left(\sqrt{\frac{2}{\pi}}\sin \theta\cos(\phi+\omega t_2)\right)\\+s_1 s_2\left(\cos ^2\tfrac{\theta}{2}~C_{ij}^{\ket0}+\sin^2\tfrac{\theta}{2}~C_{ij}^{\ket1}\right)\Bigg].
	\end{multline}
	Without loss of generality, we set $t_1=0$, and in Fig.~\ref{fig:lg2frac}, we present the full parameter-space for the two-state superpositions, showing where the $\LG{3}$ and $\LG{2}$ inequalities are violated.  Hence with superposition states, we can reach regimes III and IV from Table \ref{tbl:reg}.
	
\begin{figure}
		\includegraphics[width=16cm]{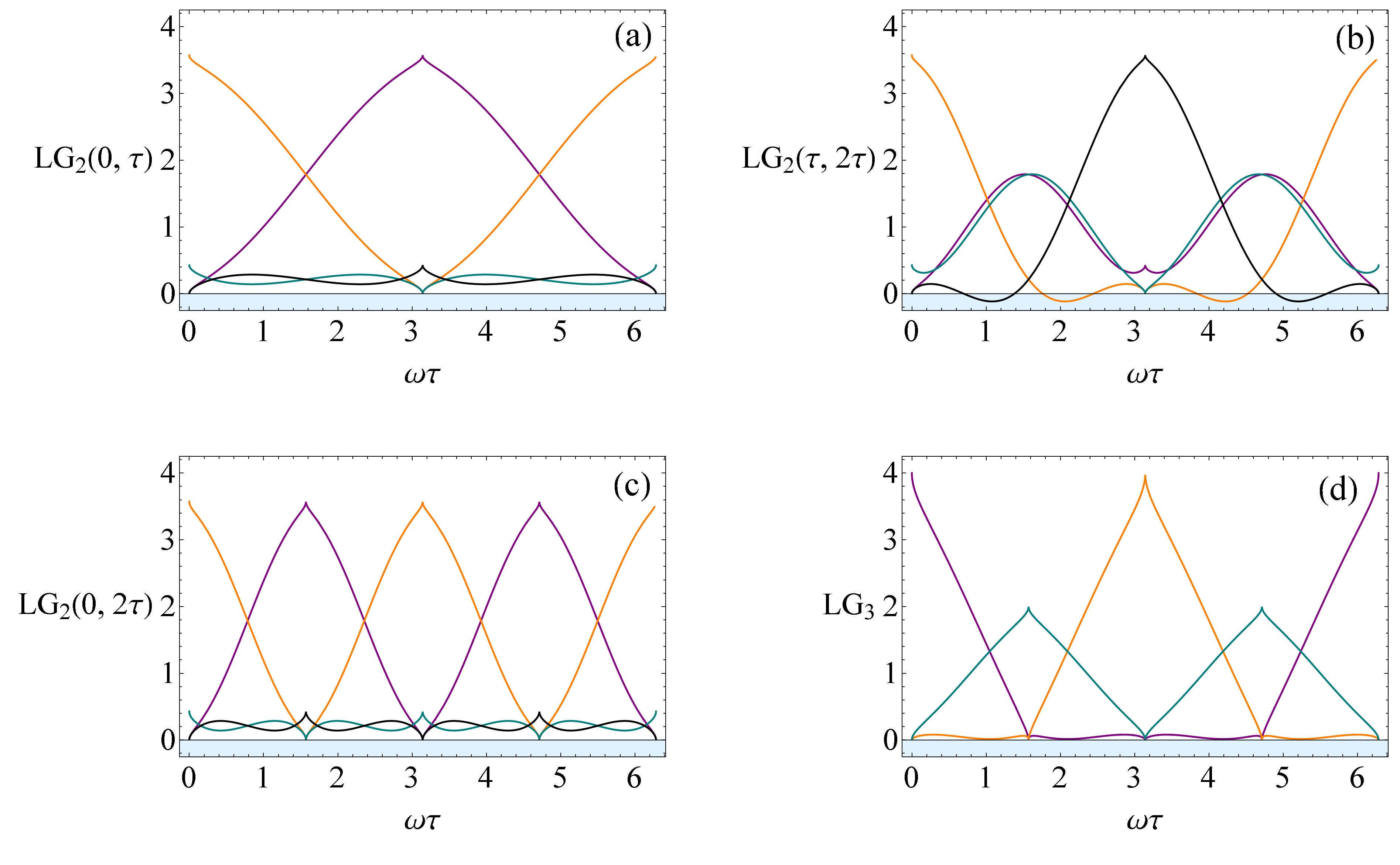}
		\caption{The complete set of two- and three-time LG inequalities are plotted for the state $\theta=1.4$, $\phi=\pi$. For this state, the $\LG{3}$ (d) inequalities are always satisfied, however there is still an LG violation present, at the level of two times, seen by the violation of the $\LG{2}$ between $\tau$ and $2\tau$ in (b).}
		\label{fig:2viol3sat}
	\end{figure}

	In Fig.~\ref{fig:2viol3sat}, the complete set of $\LG{2}$ and $\LG{3}$ inequalities are plotted for the state $\theta=0.7$, and $\phi=\pi$.  This state has been chosen to represent the important regime III, where the only violation occurs in the $\LG{2}$ inequalities between $t_2$ and $t_3$, despite the LG3s being satisfied.
	
	\subsection{LG4s and higher}
		\begin{figure}
		\subfloat[]{{\includegraphics[height=5.4cm]{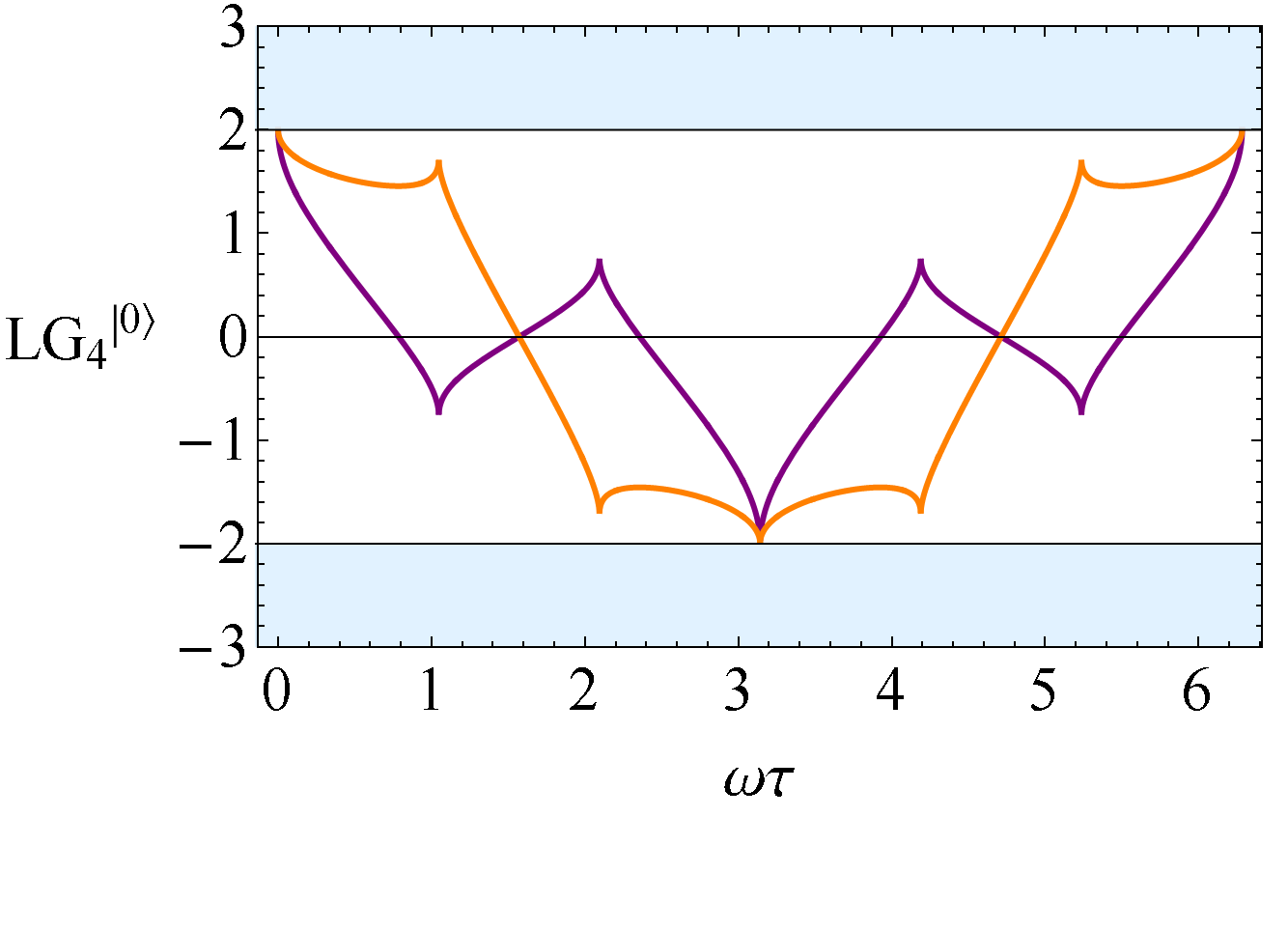}}}%
		\qquad
		\subfloat[]{{\includegraphics[height=5.4cm]{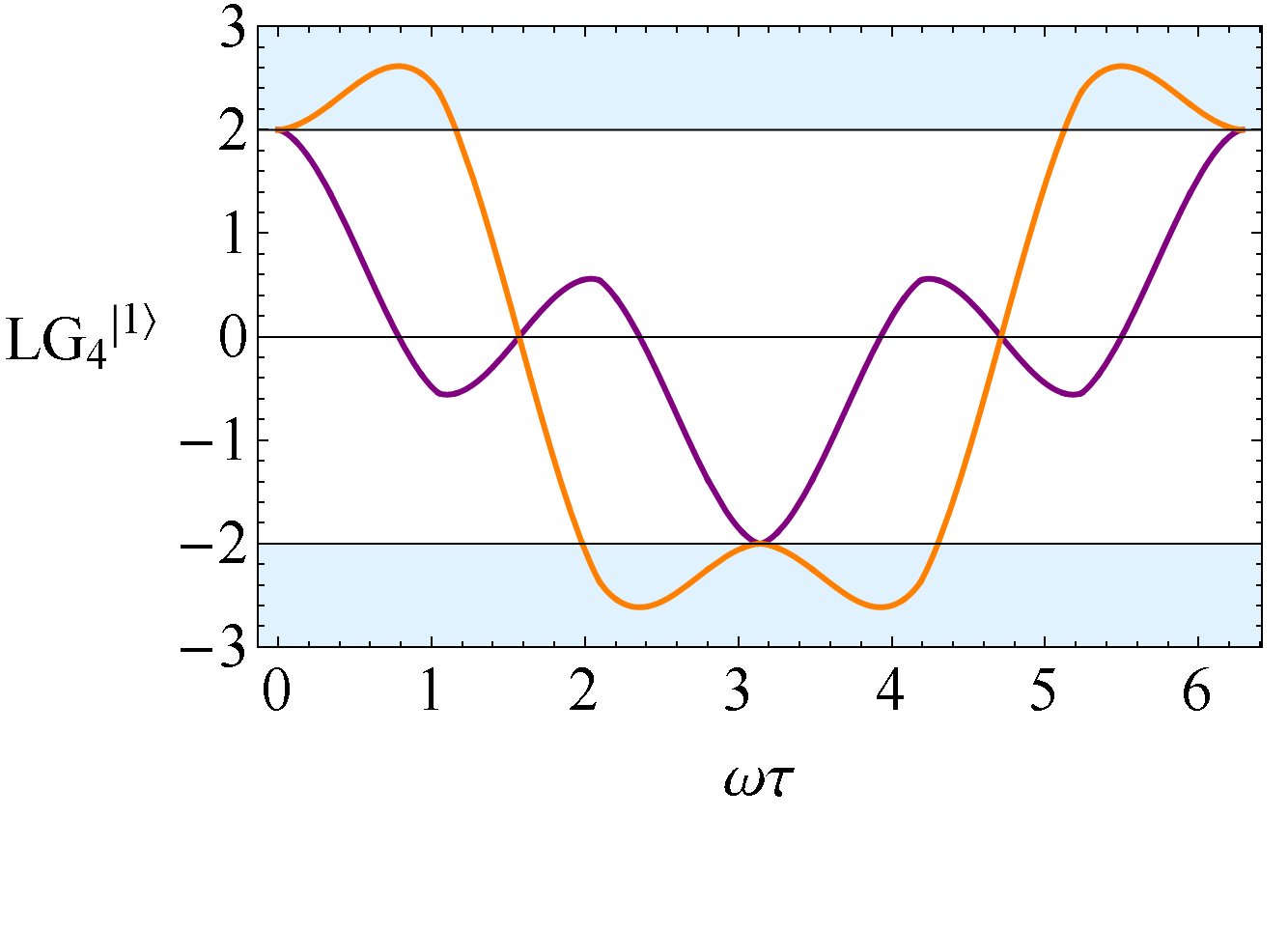}}}
		\caption{In (a) the $\LG{4}$ inequalities are plotted for the ground state, where they are satisfied for all times.  In (b), the inequalities are plotted for the first excited state, showing some regions of violation.  The maximal violation is approximately $2.615$, which represents around $92\%$ of the maximal violation of $2\sqrt{2}$.}%
		\label{fig:lg4}%
	\end{figure}
	We now consider the behaviour of the LG inequalities when more measurements are made.  We consider the case of the $\LG{4}$ inequalities, and higher order $n$-time LG inequalities \cite{halliwell2019}.  Constructed purely from sums of correlators, it is apparent that the $\LG{n}$ for two-state superpositions  will again just be mixtures of the $\ket0 $ and $\ket 1$ cases, and for $n>2$, we have 	
	\beq
	L_i(n)=\cos^2\tfrac{\theta}{2}~ L_i^{\ket0}(n)+\sin^2\tfrac{\theta}{2}~ L_i^{\ket 1}(n).
	\eeq
	The $\LG{4}$ inequalities take the form 
	\begin{equation}
	-2\leq C_{12}+C_{23}+C_{34}-C_{14}\leq 2,
	\end{equation}
	together with the six more inequalities obtained by moving the minus sign to the other three locations.  The necessary and sufficient conditions for MR at four times consist of these eight LG4 inequalities, together with the set of sixteen LG2s for the four time pairs \cite{halliwell2016b, halliwell2019, halliwell2017,halliwell2019b}.  We plot the $\LG{4}$ inequalities in  Fig.~\ref{fig:lg4}.  We again have the property that the LG4 inequalities are always satisfied for the ground state, and are violated for the first excited state.  However, in contrast to the $\LG{3}$ inequalities, the $\LG{4}$ inequalities are not violated everywhere, and have large regions of parameter space where they are satisfied.  They allow access to interesting combinations of MR violations, whilst maintaining the experimental simplicity of working with a single energy eigenstate.
	
	For the LGn case, we found no violations for the groundstate. Since the excited state correlator Eq.(\ref{eq:cos}) is very similar to the simple spin case, the asymptotic behaviour for large $n$ is the same as that of the example analyzed in our earlier paper \cite{halliwell2019}.
	
\subsection{Smoothed Projectors}
	A natural question that arises is whether the observed LG violations are an artefact of using sharp projectors, that may fade when physically realisable measurements are used.  To provide an indication that this is not the case, in Appendix \ref{sec:smooth} we repeat the calculation of $C_{12}^{\ket1}$, but with a spatially smoothed projector.  We find violations persist while the projector smoothing is less characteristic length-scale of the oscillator.
	
	\section{Higher states}\label{sec:higher}
		\begin{figure}[b]
		\subfloat[]{{\includegraphics[height=4.6cm]{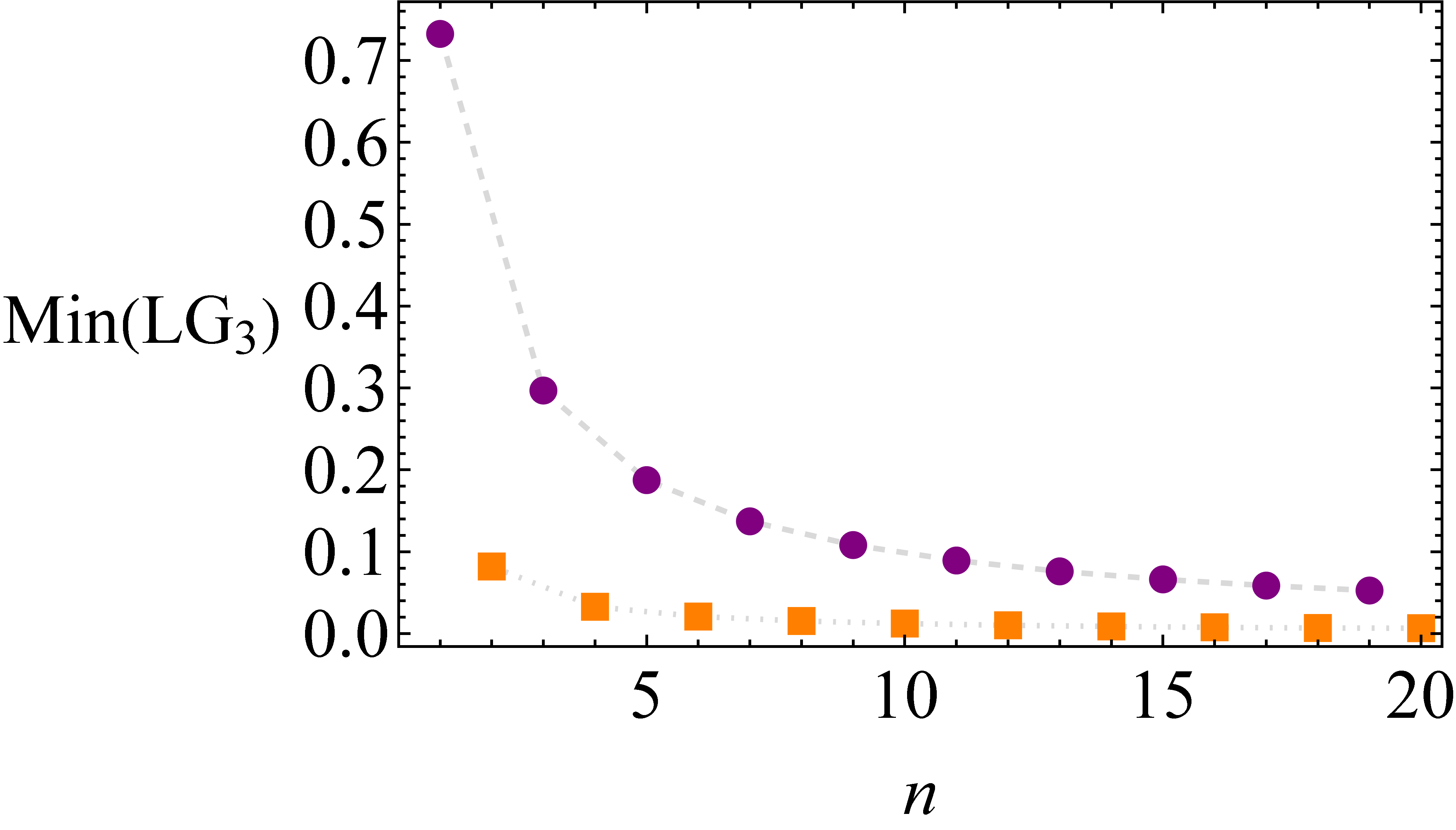}}}%
		\hspace{5mm}
		\subfloat[]{{\includegraphics[height=4.6cm]{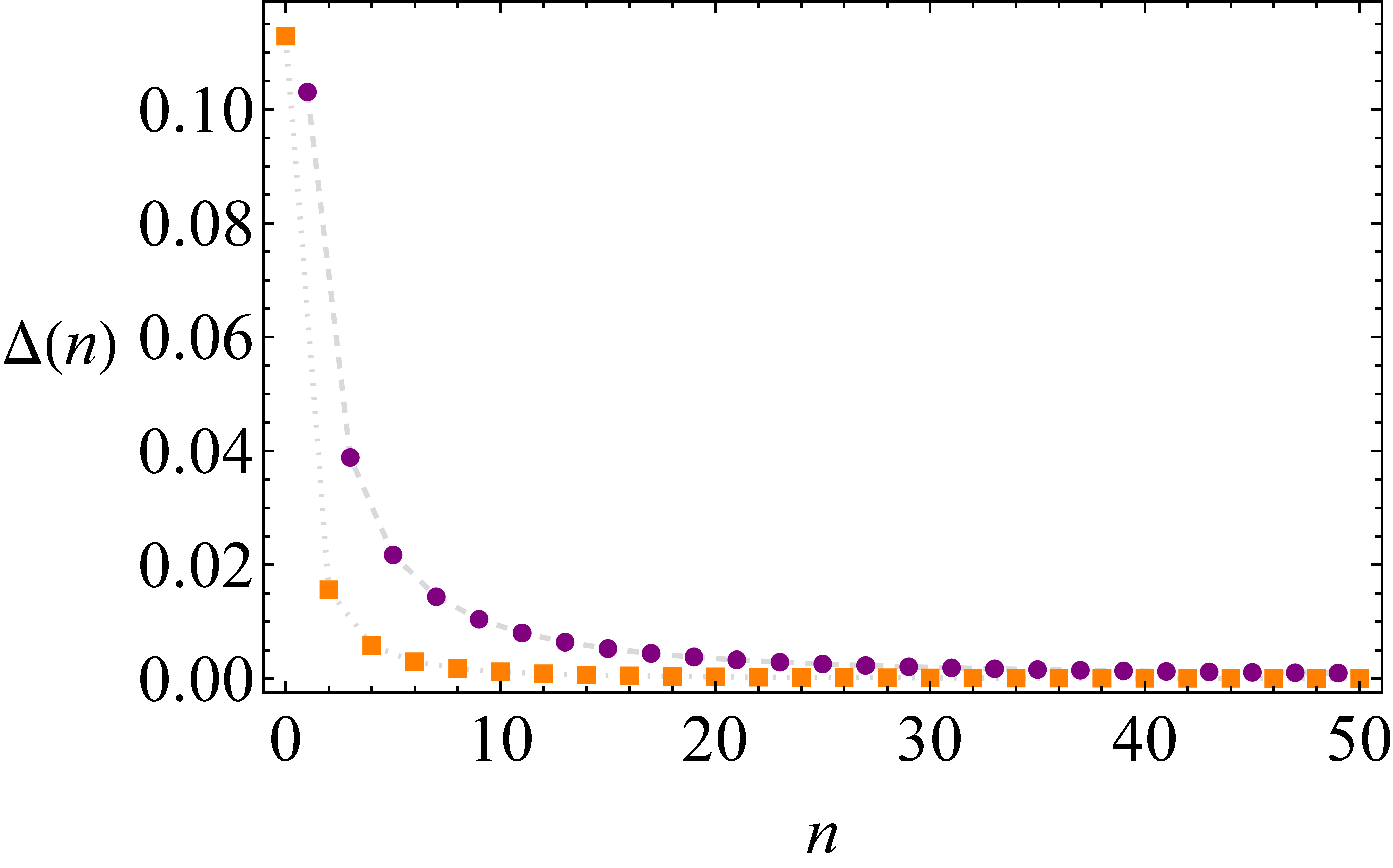}}}
		\caption{In (a), the maximal violation for a given energy eigenstate is plotted as a fraction of the L{\"u}ders bound.  The odd eigenstates are represented by the dashed line (circles), and always have significantly more violation than the even states represented by the dotted line (squares).  In (b), we plot the average magnitude of difference between the quantum correlators and the classical correlator, showing expected classicalization for large $n$. }%
		\label{fig:numopt}%
	\end{figure}
	Following Section \ref{sec:1and0}, it is natural to ask whether the LG inequalities behave significantly differently when including more of the energy eigenstate spectrum.  By two methods, we establish that for large $n$, the predictions of the QHO tend towards classical statistics.

	Firstly, by plotting the higher eigenstate correlators, we can visually see that they tend toward the classical correlator Eq.(\ref{eq:class}).  To make this exact, we calculate the average distance from the classical correlator, as a function of $n$.  For this, we define
	\begin{equation}
	\Delta(n)=\frac{1}{2\pi}\int_0^{\frac{2\pi}{\omega}}\mathop{dt}\lvert \mathbbm{C}_{12} - C_{12}^{\ket n}\rvert,
	\end{equation}
	which we plot up to $n=50$ in Fig.~\ref{fig:numopt}, where we see that the quantum temporal correlators very rapidly match the classical correlator.
	
	Secondly, with details in Appendix \ref{app:classic}, we perform an asymptotic analysis on the infinite sum representation of the quasi-probability, and show that for large $n$, this series tends towards the Fourier series of a triangle wave, matching the classical result.

	\section{Different choices of Dichotomic Variables}\label{sec:mlvl}

	We have so far worked with the simplest choice of dichotomic variable, $Q=\sgn(x)$.  We now consider what new effects may be discovered using different choices. We first consider a $Q$ defined
	using coarse grainings over more general spatial regions. Secondly, we consider a more novel choice of $Q$, namely the parity operator, which does not necessarily have a macrorealistic description but reveals some interesting features.
	
	\subsection{Quasi-probability for arbitrary regions in the ground state}
	
	We now demonstrate how to calculate the quasi-probability where each measurement is taken over an arbitrary coarse graining of space.  In this work, we consider just a pair of dichotomic variables, however the techniques presented are readily applied to a full many-valued variable LG analysis as introduced in Ref. \cite{halliwell2020}.
	
	We define projectors over arbitrary regions of space as 
	\beq
	\label{eq:genproj}
	E(\alpha)=\int_{\Delta(\alpha)}\mathop{dx}\dyad{x},
	\eeq
	where $\Delta(\alpha)$, where $\alpha=\pm$, is a set of intervals which partition the real line, which may be chosen to be different for each measurement time. The dichotomic variable $Q$ is then defined by $Q = 2 E(+) - 1 $, and discerns whether a particle is within a given region of space ($+$), or outside of it ($-$).  The two-time quasi-probability is then given by 
	\beq
	\label{eq:mlvl}
	q(+, +) =  \Re\ev{E_2(+) E_1(+)}
	\eeq
	and the LG2 inequalities are then simply $q(+,+) \ge 0 $. 
	
	Remarkably, by using these more general dichotomic variables we can find $\LG{2}$ violations in the ground state of the QHO.
	To demonstrate this, we write the quasi-probability using the QHO propagator Eq.(\ref{eq:qmnfull}), with the projectors Eq.(\ref{eq:genproj}) for the special case of the ground state, 
	\begin{equation}
	\label{eq:shortspace}
	q(+,+)=\Re\mathcal{N}_{00}(\tau)\int^{b}_a\int^{d}_c \mathop{dr}\mathop{ds}e^{-\frac{r^2}{2}}e^{-\frac{s^2}{2}}\exp({i \frac{1}{2\tan(\omega \tau)}(r^2+s^2)- i\frac{1}{\sin(\omega \tau)}rs}).
	\end{equation}
	Since the real part of the integrand oscillates around zero, negative values of $q(+,+)$ can clearly by achieved by suitable choice of the spatial intervals $[a,b]$ and $[c,d]$.
	
\begin{figure}
		\subfloat[]{{\includegraphics[height=5.8cm]{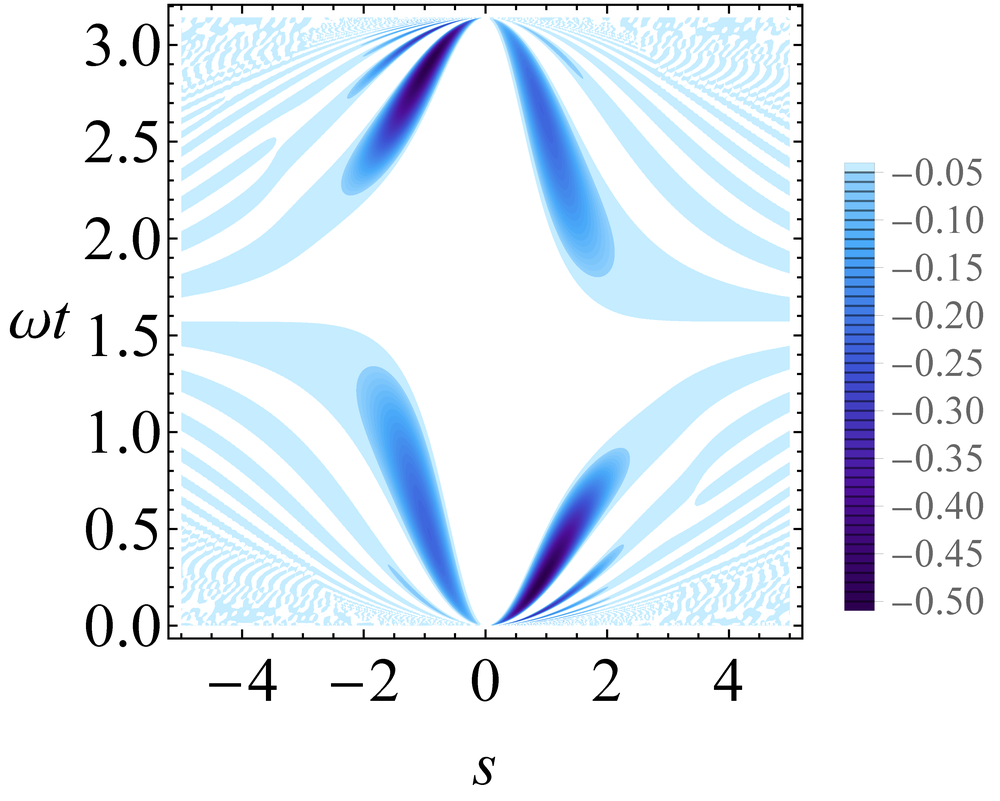}}}%
		\subfloat[]{{\includegraphics[height=5.8cm]{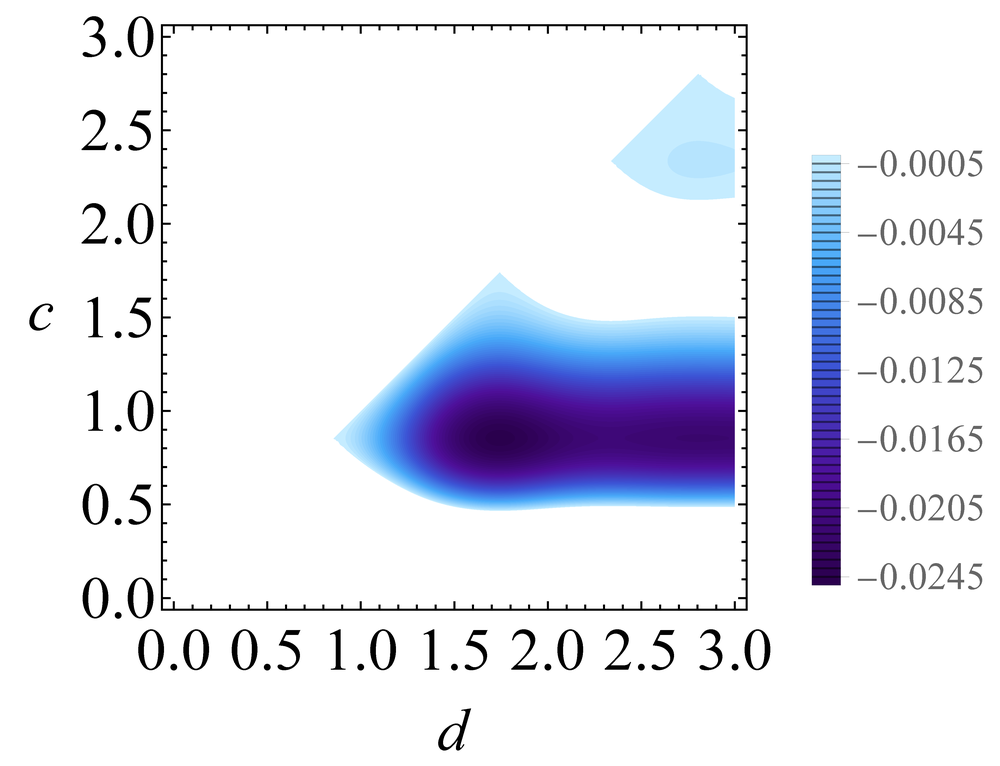}}}
		\caption{In (a), we plot the result of the first integration in Eq.(\ref{eq:shortspace}) over $[0,\infty]$, which serves as a rough map to choosing a second measurement which will lead to a negative quasiprobability.  In (b), we plot the quasi-probability, for $\omega t=2.77$, with the first measurement over $[0,\infty]$, and the second over the interval $[c,d]$.  The quasi-probability reaches around $20\%$ of the maximal violation of $-0.125$.}
		\label{fig:mlvl}
	\end{figure}

	In Fig.~\ref{fig:mlvl}(a), we plot the result of
	the integration, with the first interval  $\Delta(+)=[0,\infty)$, and then plot the remaining integrand of the second integral, in the regions where it is negative.  This indicates the region in which to make a second measurement, which will lead to negativity of the quasi-probability, and hence $\LG{2}$ violations.
	
	To calculate the quasi-probability with arbitrary coarse-grainings, it is most efficient to use the techniques in Sec.\ref{subsec:inf} since this approach already encapsulates these more general projectors.  The details of this calculation are found in Appendix \ref{app:arbproj}.
	
	In Fig.~\ref{fig:mlvl}(b), we plot the quasi-probability for $\omega \tau=2.77$, over a range of possible second measurement windows.  From this, we observe that there are a wide range of choices for the second measurement which lead to an LG2 violation, which can reach $20 \%$ of the maximum possible violation.
	We also observe that these measurements do not not need to be over particularly small regions of space, and that simply shifting the axis of the second measurement from $[0,\infty)$ to $[c,\infty)$, compared to the usual choice of dichotomic variable, is sufficient for a good LG2 violation in the ground state.	
	
	The ground state wave function has positive Wigner function and from a macrorealistic point of view, is sometimes thought of as a ``classical'' state describing a particle localized at a phase space point, yet we see here that a LG2 violation is possible. A similar phenomenon was noted in Ref.\cite{halliwell2021}, which used a coherent state with $\langle x \rangle =0$. The origin of this effect will be explored in a future paper. We have also looked for LG3 and LG4 violations in the ground state but an extensive parameter search yields no result.

	
	
	\subsection{LG violations using the parity operator}
	
	Our second choice of different dichotomic operator arises from the observation that the variables $Q_1$, $Q_2$ etc used in LG inequalities do not necessarily have to be the time evolution of a given dichotomic variable $Q$, but could be a set of {\it any} dichotomic variables. The derivation of the LG inequalities still holds so an LG test is still possible as long as the variables can be measured.
	Such an approach has been used in some experiments \cite{goggin2011}. Note at this point that we are now departing from the quantum harmonic oscillator, at least in terms of time evolution, as the physical system of interest (although the QHO is still the natural arena in which to create the gaussian state used below).
	
	Two interesting variables to use in this context are the parity operator,
	\begin{equation}
	\Pi = \int_{-\infty}^\infty dx \ | x \rangle \langle- x |,
	\end{equation}
	and the parity inversion operator,
	\begin{equation}
	R = i\int_{-\infty}^\infty dx \ {\rm sgn} (x) \ | x \rangle \langle -x |.
	\end{equation}
	These two operators, taken together with the usual choice $Q = {\rm sgn} ( x) $ have the interesting property that they obey the same algebra as the Pauli matrices
	$ [Q, R ] = 2 i  \Pi $ etc and they also all anticommute with each other. By comparison with the spin-$\frac{1}{2}$ model, this is a clear suggestion that LG violations are readily possible with such variables. These variables were previously used in 
	Ref. \cite{praxmeyer2005} to demonstrate a Bell inequality violation for a gaussian two-particle entangled state of the Einstein-Podolsky-Rosen type.
	
	We consider the LG2
	\begin{equation}
	1 - \langle Q_1 \rangle - \langle Q_2 \rangle + C_{12} \ge 0,
	\end{equation}
	and we take $Q_1 = Q = {\rm sgn} (\hat x) $ and $Q_2 = \Pi$. Since these operators anticommute the correlator is zero and the LG2 is then
	\begin{equation}
	1 - \langle Q \rangle - \langle \Pi \rangle \ge 0.
	\end{equation}
	We take a gaussian state of width $\sigma$, $\langle x \rangle = q $ and $\langle p \rangle = 0$ and readily find that the LG2 inequality is
	\begin{equation}
	1 - {\rm erf} \left( \frac{q}{\sqrt{2} \sigma } \right)  - \exp( -\frac{q^2}{2 \sigma^2})  \ge 0.
	\end{equation}
	The left-hand side is readily shown to take its minimum value of approximately $ -0.3024 $ at $ q/\sigma = \sqrt{2/\pi}$, hence there is a clear LG2 violation.
	
	Although the parity operator is measurable for some systems and therefore the above inequality can be tested experimentally, it has no macrorealistic counterpart (unlike Q = $\textrm{sgn}(x)$). Still, the fact that LG2 violations are obtained so easily with these variables could give interesting clues to understand violations in more physical cases
	
	\section{Leggett-Garg violations in the Morse potential}
	\label{sec:morse}
	In Sec. \ref{subsec:inf}, we laid out a general technique to calculate temporal correlators, when the energy eigenspectrum is known.  As a demonstration, we now apply this result to a different exactly soluble potential, the Morse potential \cite{morse1929}, an asymmetric potential with minimum at $r_c$, which combines a short-range repulsion, with a long-range attraction.  The potential is defined by
	\begin{equation}
	V(r) = D_e\left(e^{-2a(r-r_e)}-2e^{-a(r-r_e)}\right),
	\end{equation}
	where $D_e$ corresponds to the well depth, and $a$ to its width.
	
	The details and results of applying our method to the Morse potential may be found in Appendix \ref{app:morse}, where we find qualitatively similar behaviour to the QHO, and observe significant $\LG{3}$ and $\LG{4}$ violations in the $\ket1$ state.  There are several other systems for which the Schr\"{o}dinger equation may be solved exactly, including the Hydrogen-like atom \cite{landau1977}, the linear potential \cite{khonina2013}, and the quantum pendulum \cite{galvez2019}.  The result presented in Sec.~\ref{subsec:inf} hence allows future exploration of the behaviour of the LG inequalities in several bound systems.

	\section{Summary and Conclusion}
	\label{sec:conc}
	We have presented an analytic investigation of the LG approach to macrorealism in bound systems with an in-depth study of the QHO.  
	For a dichotomic variable given by the sign of the oscillator position,
	we developed a method to calculate the temporal correlators for energy eigenstates of any bound system with a known energy eigenspectrum.
	
	We found that all the interesting variants of MR conditions tested for by combinations of the \LG{2}, \LG{3} and \LG{4} inequalities can be observed in states of the QHO involving superpositions of only the $\ket0$ and $\ket1$ states.  
	Intriguingly, the temporal correlator of the $\ket1$ state was shown to be, to a very good approximation, a cosine, like the temporal correlators in the canonical simple spin systems used in much LG research.  	We also found that the higher energy eigenstates rapidly begin to exhibit statistics in line with a classical model.
	
	We found that LG violations persist even with a significant amount of smoothing of the projectors.  This is important, since it means the LG violations are not just an artefact of non-physical, sharp projectors, and therefore the LG violations described here should be directly observable in experiment.
	
	Although the bulk of the work in this paper has been conducted using the simplest dichotomic variable $ Q =\text{sgn}(x)$, all the results presented are general enough to also allow for dichotomic variables defined on arbitrary regions of space.  We briefly investigated the effect of using a different variable, and found it provides a richer and more powerful test of MR, with two-time LG violations even in the $\ket0$ state, despite being a state with positive Wigner function. (See also Ref. \cite{halliwell2021}).  In addition, we found LG2 violations if one of the dichotomic variables at one of times is taken to be the parity operator.
	
	To demonstrate the method, we calculated the temporal correlators for another exactly soluble system, the Morse potential, where we found significant LG3 and LG4 violations for the first excited state.
	
	In conclusion, we have derived a number of results on LG inequalities for coarse-grained position measurements in bound systems. These results provide indications as to what sort of states to 
	create in experiments on macroscopic systems in order to find evidence of macroscopic coherence.
	
	Future work will explore in more depth the phenomenon of LG violations in these systems for initial states with positive Wigner function. We also propose to develop non-invasive measurement protocols suitable for continuous variable systems by adapting the continuous in time velocity measurement scheme proposed and utilized in Refs. \cite{halliwell2016a,majidy2019a}.

	\section*{Acknowledgements}
	
	We are grateful to Sougato Bose and Dipankar Home for useful discussions.  We also thank Michael Vanner for discussions on the experimental feasibility of the tests discussed in this paper. CM thanks Prof.~Adams for choosing to include the rather arcane Node theorem within MIT OCW’s free Quantum Physics lectures, which many years ago supplemented my learning and helped to build my foundations in the subject.

	%
	
	\appendix

	\section{Quasi-probability for generic bound potentials}
	\label{app:general}
	\subsection{Partial overlap integrals}
	\label{app:wronski}
	To calculate the quasi-probability most generally, we need the integrals
	\begin{equation}
	J_{k\ell}(x_1, x_2)=\int_{x_1}^{x_2}\mathop{dx}\braket{k}{x}\braket{x}{\ell},
	\end{equation}
	where we adopt the notation that when $J_{k\ell}$ appears without an argument, $J_{k\ell}=J_{k\ell}(0,\infty)$. It is possible to compute these integrals, by expanding a result found in Ref. \cite{moriconi2007a}, which bears resemblance to Abel's identity.  We first denote energy eigenstates in position space as
	\begin{equation}
	\psi_n(x) = \braket{x}{n}.
	\end{equation}
	We then construct the Wronskian between the two states $k$ and $\ell$,
	\begin{equation}
	W(x)=\psi_k'\psi_\ell - \psi_\ell'\psi_k.
	\end{equation}
	Differentiating the Wronskian, we find
	\begin{equation}
	\label{eq:wdiff}
	W'(x)=\psi_k''\psi_\ell - \psi_\ell'' \psi_k.
	\end{equation}
	Since we are working with energy eigenstates, from the Schr{\"o}dinger equation we have
	\begin{equation}
	\label{eq:schrono}
	\psi_n'' =2(V-\varepsilon_n)\psi_n,
	\end{equation}
	where $\varepsilon_n$ are the dimensionless energy eigenvalues. Substituting this into Eq.(\ref{eq:wdiff}), we find
	\begin{equation}
	W'(x)=2(\varepsilon_\ell-\varepsilon_k)\psi_k\psi_\ell,
	\end{equation}
	where any dependence on the form of the potential has disappeared from this equation, being contained implicitly in the spectrum of the Hamiltonian.  We are now free to integrate both sides over any given region of space, which yields
	\begin{equation}
	\frac{1}{2(\varepsilon_\ell-\varepsilon_k)}\eval{W(x)}_{x_1}^{x_2}=\int_{x_1}^{x_2}\mathop{dx}\psi_k\psi_\ell.
	\end{equation}
	The right-hand side of the equation is exactly the matrix elements $J_{k\ell}$, and so completing the integration, we find
	\begin{equation}
	\label{eqn:wronskiapp}
	J_{k\ell}=\frac{1}{2(\varepsilon_\ell- \varepsilon_k)}\left[\psi_k'(x_2)\psi_\ell(x_2)-\psi_\ell'(x_2)\psi_k(x_2)-\psi_k'(x_1)\psi_\ell(x_1)+\psi_\ell'(x_1)\psi_k(x_1)\right].
	\end{equation}
	\subsection{Calculational Details}
	\label{app:calcdeets}
	
	We continue the analysis coarse-graining onto the left and right hand side of the axis, in the important special case of symmetric potentials.  In this case, we have
	\begin{equation}
	\label{eq:jijlim}
	J_{ij}=\int_{0}^{\infty}\mathop{dx}\psi_i\psi_j=\frac{1}{2(\varepsilon_j-\varepsilon_i)}\left[\psi_j'(0)\psi_i(0)-\psi'_i(0)\psi_j(0)\right],
	\end{equation}
	since wavefunctions must vanish at infinity.  With a symmetric potential, we have for $n$ odd, $\psi_n(0)=0$, and for $n$ even, we have $\psi_n'(0)=0$.  Eq.(\ref{eq:wdiff}) is not defined for $i=j$, however for a symmetric potential, we know $J_{ii} = \frac{1}{2}$, as so the $k=n$ term contributes $\frac14$ is to both results.
	
	Hence, if $n$ is odd, we have
	\begin{equation}
	\label{eq:qgen}
	q(n)=\frac14+\Re e^{i \frac{E_n}{\hbar} \tau}\psi_n'(0)^2\sum_{k=0, k\neq n}^{\infty}e^{-i \frac{E_k}{\hbar} \tau}\frac{1}{4(\varepsilon_k-\varepsilon_n)^2}\psi_k(0)\psi_k(0).
	\end{equation}
	If $n$ is even, we have
	\begin{equation}
	q(n)=\frac14+\Re e^{i\frac{E_n}{\hbar} \tau}\psi_n(0)^2\sum_{k=0,k\neq n}^{\infty}e^{-i \frac{E_k}{\hbar} \tau}\frac{1}{4(\varepsilon_k-\varepsilon_n)^2}\psi_k'(0)\psi_k'(0).
	\end{equation}
	
	To be computationally feasible, we must truncate this sum for some $m$.  To estimate the error involved with a given truncation, we note that $J_{k\ell}$ form the coefficients of the expansion of $\theta(\hat x)\ket{k}$. This represents the probability of finding the particle on either side of the axis.  In the case of symmetric potentials, these are  (anti-)symmetric states, and so have norm $\frac12$.  Hence this gives us the truncation 
	\beq
	\label{eq:trunc}
	\Delta_n(m)=\frac12-
	\sum_{k=0}^{m}J_{nk}^2.
	\eeq
	\section{Exact Quasiprobability in the QHO}
	\label{app:gen}
	\subsection{Strategy}
	\label{app:strat}
	To calculate the quasi-probability exactly, we begin by writing Eq.(\ref{eq:qmnfull}) as 
	\beq
	q_{mn}=\mathcal{N}e^{i\frac{E_m}{\hbar}t_2-i\frac{E_n}{\hbar}t_1}\int^{\infty}_0\int^{\infty}_0 \mathop{dr}\mathop{ds}H_m(r)H_n(s)\exp(\alpha r^2 + \beta s^2+ i \gamma r s),
	\eeq
	where we have introduced the variables
	\begin{align}
	\alpha = \beta &= \frac{i}{2 \tan(\omega \tau)}-\frac{1}{2},\\
	\gamma &= -\frac{1}{\sin(\omega \tau)}.
	\end{align}
	We now note that as a product of Hermite polynomials, the term $H_m(r)H_n(s)$ will itself just be a polynomial involving powers of $r$ and $s$. Hence $q_{mn}$ may be broken up into a sum of terms of the form
	\beq
	X_{k\ell}=\int_0^{\infty}\int_0^{\infty}\mathop{dr}\mathop{ds}r^k s^\ell \exp(\alpha r^2 +\beta s^2 + i\gamma r s).
	\eeq
	To calculate these terms, we consider the generating integrals
	\begin{align}
	\label{genI}
	I(\alpha, \beta,\gamma)&=\int^{\infty}_0\int^{\infty}_0\mathop{dr}\mathop{ds}\exp(\alpha r^2+\beta s^2 + i\gamma r s),\\
	\label{genJ}
	J(\alpha, \beta,\gamma)&=\int^{\infty}_0\int^{\infty}_0\mathop{dr}\mathop{ds}r\exp(\alpha r^2+\beta s^2 + i\gamma r s).
	\end{align}
	We may then calculate any of the $X_{k\ell}$ integrals through repeated use of partial differentiation.  In particular, if $k$ and $\ell$ are both even, we have
	\beq
	X_{k\ell}=\pdv{\alpha^{\frac{k}{2}}}\pdv{\beta^{\frac{\ell}{2}}}I(\alpha,\beta,\gamma).
	\eeq
	Similarly, if $k$ and $\ell$ are both odd, we have 
	\beq
	X_{k\ell}=-i\pdv{\gamma}\pdv{\alpha^{\frac{k-1}{2}}}\pdv{\beta^{\frac{\ell-1}{2}}}I(\alpha,\beta,\gamma).
	\eeq
	Finally, for the case where say $k$ is odd, and $\ell$ is even, we have
	\beq
	X_{k\ell}=\pdv{\alpha^{\frac{k-1}{2}}}\pdv{\beta^{\frac{\ell}{2}}}J(\alpha, \beta, \gamma),
	\eeq
	and vice-versa for $\ell$ odd, $k$ even.  This approach although a handful on paper, is simple to implement in computer algebra software.
	\subsection{Generating Integrals}
	We now proceed with calculating $I(\alpha,\beta,\gamma)$ Eq.(\ref{genI}), by completing the square in the exponential function.  This yields
	\begin{equation}
	I(\alpha,\beta,\gamma)=\int_0^\infty \int_0^\infty \mathop{dr}\mathop{ds}\exp\left(\frac{r^2(4\alpha \beta + \gamma^2)}{4\beta}\right)\exp\left(\beta s+i \frac{\gamma r}{2\sqrt{\beta}}\right).
	\end{equation}
	We introduce the short-hand $\delta=\frac{4\alpha \beta + \gamma^2}{4\beta}$.  The $s$ integral may now be completed in terms of the error function as
	\begin{equation}
	\label{eq:iint}
	I(\alpha,\beta,\gamma)=\frac{\sqrt{\pi}}{2\sqrt{-\beta}}\int_0^\infty \mathop{dr}\exp\left(\delta r^2\right)\left(1+i\erfi\left(\frac{\gamma r}{2\sqrt{-\beta}}\right)\right).
	\end{equation}
	For simplicity of presentation, we separate the integral into two parts,
	\begin{align}
	I_1(\alpha, \beta,\gamma)&=\frac{\sqrt{\pi}}{2\sqrt{-\beta}}\int_0^\infty \mathop{dr}\exp\left(\delta r^2\right),\\
	I_2(\alpha, \beta,\gamma)&=\frac{i\sqrt{\pi}}{2\sqrt{-\beta}}\int_0^\infty \mathop{dr}\exp\left(\delta r^2\right)\erfi\left(\frac{\gamma r}{2\sqrt{-\beta}}\right).
	\end{align}
	The first integral is simply the Gaussian integral on the half-plane, which has the result
	\begin{equation}
	I_1(\alpha, \beta,\gamma)=\frac{\pi}{4\sqrt{-\beta}}\sqrt{-\frac{1}{\delta}}.
	\end{equation}
	To proceed with $I_2(\alpha, \beta,\gamma)$, we rescale the variables to $u=\frac{\gamma r}{2\sqrt{-\beta}}$, leading to
	\begin{equation}
	I_2(\alpha,\beta,\gamma)=\frac{i\sqrt{\pi}}{\gamma}\int_0^{\infty}\mathop{du}\exp(-u^2\frac{4\alpha\beta+\gamma^2}{\gamma^2})\erf(u).
	\end{equation}
	This integral takes the form
	\begin{equation}
	\int_0^\infty \mathop{du}e^{-c u ^2}\erf{u}=\frac{1}{\sqrt{\pi}\sqrt{c}}\arctan\left(\frac{1}{\sqrt{c}}\right),
	\end{equation}
	which is convergent for $\Re(c)>0$, which may be confirmed for the case with $c=\frac{4\alpha \beta +\gamma^2}{\gamma^2}$.  Overall, this gives the result 
	\begin{equation}
	I_2(\alpha,\beta,\gamma)=\frac{1}{\sqrt{4\alpha \beta + \gamma^2}}\arctan\left(\frac{i\gamma}{\sqrt{4\alpha \beta + \gamma^2}}\right).
	\end{equation}
	Hence the complete result is
	\begin{equation}
	I(\alpha,\beta,\gamma)=\frac{\pi}{4\sqrt{-\beta}}\sqrt{-\frac{1}{\delta}}\left(1+\frac{2}{\pi}\arctan\left(\frac{i\gamma}{\sqrt{4\alpha \beta + \gamma^2}}\right)\right).
	\end{equation}
	
	To calculate $J(\alpha, \beta,\gamma)$, we pick up the calculation from Eq.(\ref{eq:iint}), with
	\beq
	J(\alpha,\beta,\gamma)=\frac{\sqrt{\pi}}{2\sqrt{-\beta}}\int_0^\infty \mathop{dr}r\exp\left(\delta r^2\right)\left(1+i\erfi\left(\frac{\gamma r}{2\sqrt{-\beta}}\right)\right),
	\eeq
	which for clarity, we again separate into two parts, 
	\begin{align}
	J_1(\alpha, \beta,\gamma)&=\frac{\sqrt{\pi}}{2\sqrt{-\beta}}\int_0^\infty \mathop{dr}r\exp\left(\delta r^2\right),\\
	J_2(\alpha, \beta,\gamma)&=\frac{i\sqrt{\pi}}{2\sqrt{-\beta}}\int_0^\infty \mathop{dr}r\exp\left(\delta r^2\right)\erfi\left(\frac{\gamma r}{2\sqrt{-\beta}}\right).
	\end{align}
	These integrals are all simpler to compute, owing to the presence of the factor of $r$.  The first integral is easily calculated using $\pdv{r}e^{ar^2}=2a r e^{ar^2}$, and is
	\beq
	J_1(\alpha,\beta,\gamma)=-\frac{\sqrt{\pi}}{4\sqrt{-\beta}}\frac{1}{\delta}.
	\eeq
	The second integral is amiable to integration by parts, where we find the result
	\beq
	J_2(\alpha,\beta,\gamma)=-\frac{i \sqrt{\pi } \gamma }{2 \beta  \sqrt{\frac{\alpha  \beta }{4 \alpha  \beta +\gamma ^2}} \left(-\frac{4 \alpha  \beta +\gamma ^2}{\beta }\right)^{3/2}}.
	\eeq
	In total, this gives the result
	\begin{equation}
	J(\alpha,\beta,\gamma)=\frac{\sqrt{\pi}}{2\sqrt{-\beta}}\left(-\frac{2\beta}{4\alpha \beta + \gamma^2}+i\frac{\gamma}{\sqrt{-\beta}\sqrt{\frac{\alpha \beta}{4\alpha \beta + \gamma^2}}\left(-\frac{4\alpha \beta + \gamma^2}{\beta}\right)^{\frac{3}{2}}}\right).
	\end{equation}
	\subsection{Example Calculations}
	With the generating integrals completed, we can find the expressions for correlators using the approach in Appendix~\ref{app:strat}.
	
	The final results are significantly simplified by the identity
	\beq
	\label{eqn:id}
	\frac{\beta}{4\alpha \beta + \gamma^2}=-\frac{1}{4},
	\eeq
	which is simple to prove, by substituting in the definitions for $\alpha,\beta $ and $\gamma$ as
	\begin{equation}
	\frac{\beta}{4\alpha \beta + \gamma^2}=\frac12 \frac{-1+i \cot \omega \tau}{1-2i \cot \omega \tau -\cot^2 \omega \tau+\csc^2\omega \tau},
	\end{equation}
	where using the identity $\csc^2x-\cot^2x=1$ yields the required result.  
	
	It is important that this identity only be applied after completing all the differentiation steps of the algorithm in Appendix~\ref{app:strat}.  
	
	We also note the use of the factor of $\frac{1}{\sqrt{-\beta}}$
	\beq
	\frac{1}{\sqrt{2i\sin\omega \tau}}=e^{-i\frac{\omega \tau}{2}}\sqrt{-\beta},
	\eeq
	which allows us to take care of the time-dependence of the propagator prefactor.
	
	To for example calculate the correlator for the ground state, we find
	\begin{equation}
	q_{00} = \frac{1}{\pi} e^{-i\frac{\omega \tau}{2}}\sqrt{-\beta}e^{i\frac{\omega t_2}{2}-i\frac{\omega t_1}{2}}I(\alpha,\beta,\gamma).
	\end{equation}
	Making the appropriate cancellations and substitutions we find
	\begin{equation}
	q_{00}=\left(\sqrt{-\frac{\beta}{4\alpha \beta + \gamma^2}}+\frac{2}{\pi}\sqrt{-\frac{\beta}{4\alpha \beta + \gamma^2}}\arctan\left(\frac{i\gamma}{\sqrt{4\alpha\beta + \gamma^2}}\right)\right).
	\end{equation}
	Making use of the identity Eq.(\ref{eqn:id}), we find
	\begin{align}
	\label{q00}
	q_{00}&=\frac{1}{4}\left(1+\frac{2}{\pi} \arctan(\frac{\gamma}{\sqrt{-4 \alpha \beta - \gamma^2}})\right),
	\end{align}
	whereupon back-substitution of the $\alpha,\beta$ and $\gamma$, and taking the real part, we find
	\begin{align}
	q^{\ket0}(+,+)=\frac14 \left(1+\frac{2}{\pi}\Re \arctan(\frac{ie^{i\frac{\omega\tau}{2}}}{\sqrt{2i\sin\omega\tau}})\right).
	\end{align}
	Care must be taken with the branch-cut of the square-root function, and the results presented here are consistent with the choice of taking the branch cut along the negative real axis.
	\subsection{Temporal Correlators}
	\label{app:corrs}
	We tabulate the exact correlators for the first nine energy eigenstates of the QHO.  We again rely on the function $f(\tau)=-i e^{-i\frac{\omega \tau}{2}}\sqrt{2i\sin\omega \tau}$.
	
	\begin{align}
	C_{ij}^{\ket0}&=\frac{2}{\pi}\Re\left[\arctan\left(\frac{1}{f(\tau)}\right)\right],\\
	C_{ij}^{\ket1}&=\frac{2}{\pi}\Re\left[\arctan\left(\frac{1}{f(\tau)}\right)+f(\tau)\right],\\
	C_{ij}^{\ket2}&=\frac{2}{\pi}\Re\left[\arctan\left(\frac{1}{f(\tau)}\right)+\frac{1}{2}f(\tau)\right],\\
	C_{ij}^{\ket3}&=\frac{2}{\pi}\Re\left[\arctan\left(\frac{1}{f(\tau)}\right)+\frac{5+e^{-2i\omega\tau}}{6}f(\tau)\right],\\
	C_{ij}^{\ket4}&=\frac{2}{\pi}\Re\left[\arctan\left(\frac{1}{f(\tau)}\right)+\frac{14+e^{-2i\omega\tau}}{24}f(\tau)\right],\\
	C_{ij}^{\ket5}&=\frac{2}{\pi}\Re\left[\arctan\left(\frac{1}{f(\tau)}\right)+\frac{94+17e^{-2i\omega\tau}+9e^{-4i\omega\tau}}{120}f(\tau)\right],\\
	C_{ij}^{\ket6}&=\frac{2}{\pi}\Re\left[\arctan\left(\frac{1}{f(\tau)}\right)+\frac{148+14e^{-2i\omega\tau}+3e^{-4i\omega\tau}}{240}f(\tau)\right],\\
	C_{ij}^{\ket7}&=\frac{2}{\pi}\Re\left[\arctan\left(\frac{1}{f(\tau)}\right)+\frac{1276+218^{-2i\omega\tau}+111e^{-4i\omega\tau}+75e^{-6i\omega \tau}}{1680}f(\tau)\right],\\
	C_{ij}^{\ket8}&=\frac{2}{\pi}\Re\left[\arctan\left(\frac{1}{f(\tau)}\right)+\frac{8528+904e^{-2i\omega\tau}+258e^{-4i\omega\tau}+75e^{-6i\omega \tau}}{13440}f(\tau)\right].
	\end{align}
	\section{QHO Results}
	\subsection{Smoothed Projectors}
	\label{sec:smooth}
	It is simplest to calculate the effect of smoothed projectors, working with the quasi-probability expressed as a truncated infinite sum.  To do this, we switch from using $\theta(\hat x)$ as our projector, to the continuous $\frac12(1+ \erf(\sqrt{\frac{m\omega}{\hbar}}\frac{\hat x}{a}))$. Where $a$ is a dimensionless parameter characterising the degree of smoothing.  For small $a$, we expect to recover the sharp projector result, and $a=1$ corresponds to a smoothing on the characteristic length scale of the QHO, $\sqrt{\frac{m\omega}{\hbar}}$.	This adjusts the matrix elements to be
	
	\beq
	\label{Jil}
	J_{k\ell}=\frac{1}{\sqrt{\pi2^{k+\ell} k!\ell!}}\int_{-\infty}^{\infty}\mathop{dr}H_{k}(r)H_{\ell}(r)e^{-r^2}\frac{1+ \erf(\frac{r}{a})}{2}.
	\eeq
	
\begin{figure}
		\subfloat[]{{\includegraphics[height=4.6cm]{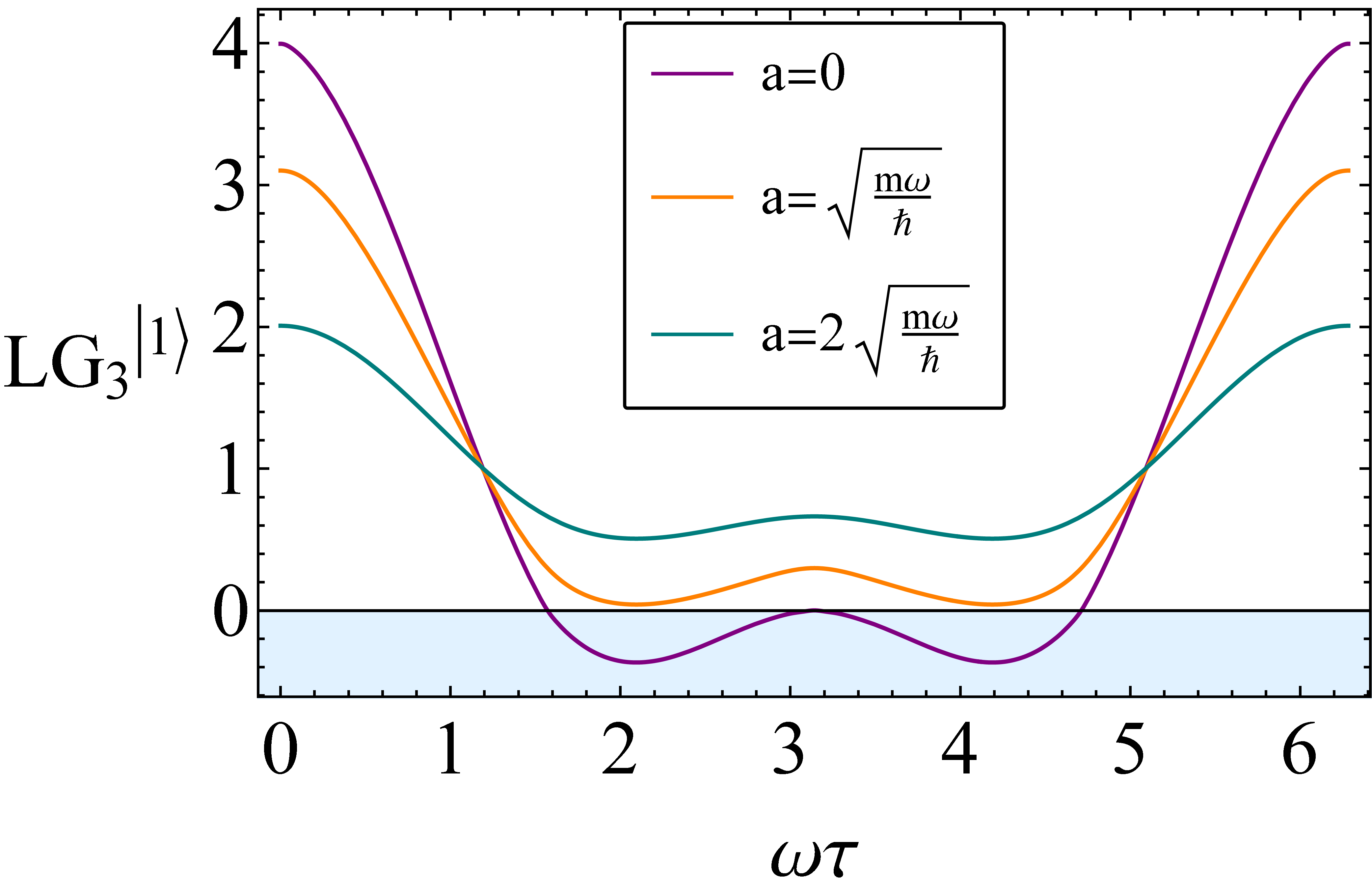}}}%
		\hspace{5mm}
		\subfloat[]{{\includegraphics[height=4.6cm]{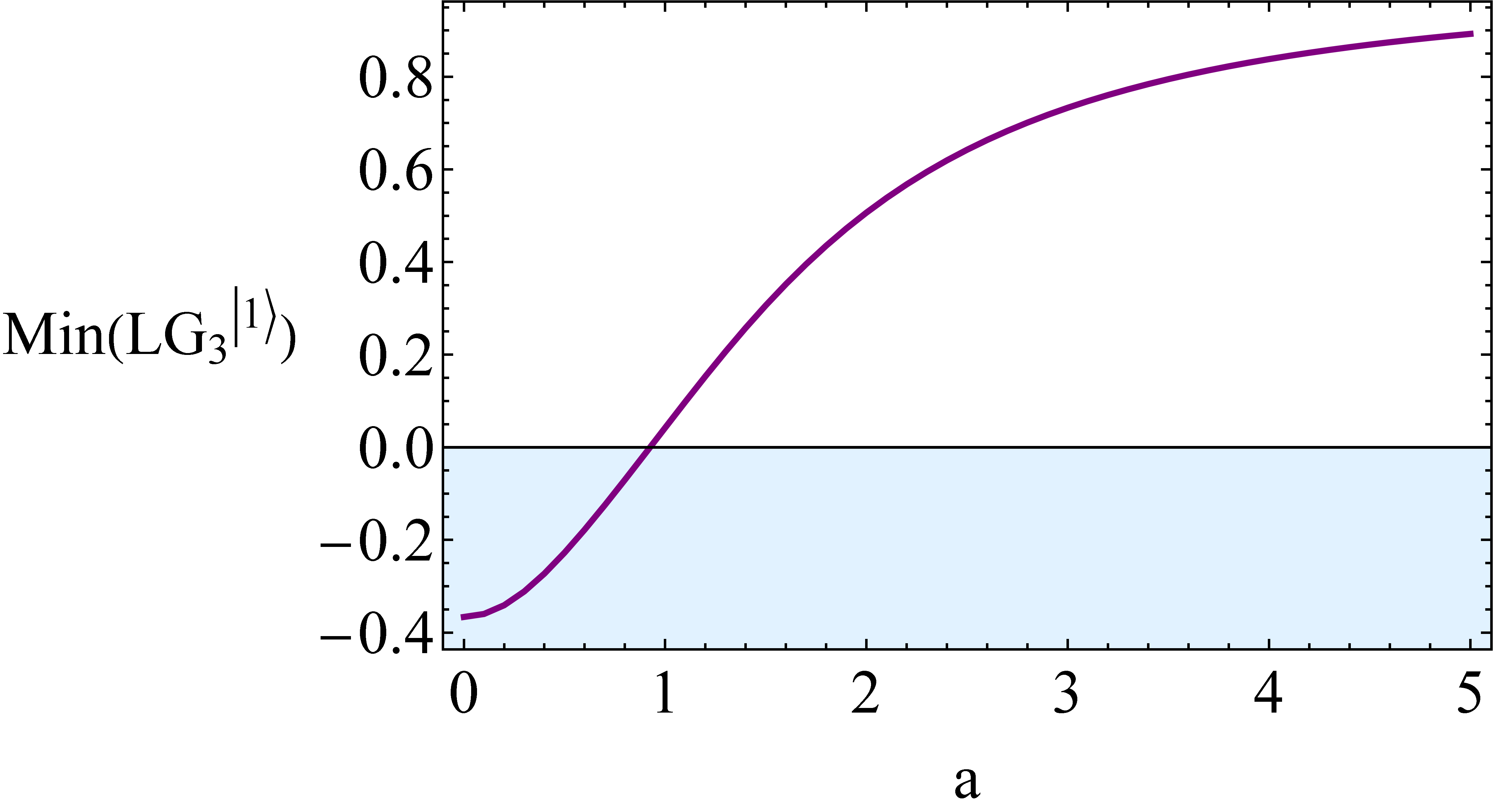}}}
		\caption{In (a), we plot one of the $\LG{3}$s, for different values of $a$, showing the qualitative effect of projector smoothing.  In (b), we plot the minimum value taken by the $\textrm{LG}_3$ inequalities, for a given value of $a$.}%
		\label{fig:smoothp}%
	\end{figure}

	We compute these integrals numerically, and then investigate the LG violations possible with different values of $a$.  In Fig.~\ref{fig:smoothp}(a), we plot one of the $\LG{3}$ inequalities, varying the value of $a$.  The smoothed projectors result in a similar (although not identical) shape, but with a reduced amplitude.  In Fig.~\ref{fig:smoothp}(b), we plot the minimal value taken by the $\LG{3}$ inequalities for each value of $a$, where we can see that once the smoothing reaches the characteristic length-scale of the QHO, LG violations vanish.
	\subsection{Classicalization}
	\label{app:classic}
	To understand the classicalization of the QHO, we look at the large $n$ asymptotic behaviour of the quasi-probability, using the original sharp projectors. Using Eq.(\ref{eq:qgen}) for the QHO, we find
	\begin{equation}
	q(n)=\frac14+\Re \psi_n'(0)^2 \sum_{k=0}^{\infty}e^{-i(k-n)\tau}\frac{1}{4(k-n)^2}\psi_k(0)^2,
	\end{equation}
	where the sum is over even $k$.  We first relabel the sum to simplify the denominator,
	\begin{equation}
	q(n)=\frac14+\Re \psi_n'(0)^2 \sum_{k=-n}^{\infty}e^{-ik\tau}\frac{1}{4k^2}\psi_{k+n}(0)^2,
	\end{equation}
	which also shifts it to a sum over only odd $k$.  We now make the first of two observations relevant in the large $n$ regime.  Since the Hermite functions evaluated at $0$ remain finite, and of order of magnitude $1$, the magnitude of the terms of this sum will fit into the envelope given by $\frac{1}{k^2}$, and so, for large $n$, we can safely extend the lower limit of the sum to $-\infty$, yielding
	\begin{equation}
	q(n)=\frac14+\Re \psi_n'(0)^2 \sum_{k=-\infty}^{\infty}e^{-ik\tau}\frac{1}{4k^2}\psi_{k+n}(0)^2.
	\end{equation}
	Using the standard recurrence relations of the Hermite functions, we find that at the origin we have
	\begin{equation}
	\frac{\psi_{n+1}(0)}{\psi_{n-1}(0)}=\sqrt{\frac{n}{n+1}}.
	\end{equation}
	This leads to the second observation, that in the large $n$ regime, the rate at which $\psi_{n+k}(0)$ changes is negligible compared to the rate of change of $\frac{1}{k^2}$, and hence may be considered approximately constant in the sum, and so we have
	\begin{equation}
	\label{eq:progress}
	q(n)\approx\frac14+\Re \psi_n'(0)^2 \psi_{n+1}(0)^2\sum_{k=-\infty}^{\infty}e^{-ik\tau}\frac{1}{4k^2}.
	\end{equation}
	To proceed, we look at the generating function of the Hermite polynomials, 
	\begin{equation}
	\label{eqn:gen}
	e^{2xt-t^2}=\sum_{n=0}^{\infty}H_n(x)\frac{t^n}{n!}
	\end{equation}
	Evaluating at $x=0$, and using the Taylor series for the Gaussian, we find
	\begin{equation}
	\sum_{k=0}^\infty (-1)^k\frac{t^{2k}}{k!}=\sum_{n=0}^{\infty}H_n(0)\frac{t^n}{n!},
	\end{equation}
	whereby comparing powers of $t$, we find for even $n$, $\abs{H_n(0)}=\frac{n!}{(\frac{n}{2})!}$.  Performing a similar analysis for the derivative term, we find that for odd $n$, $\abs{H'_n(0)}=\frac{(n+1)!}{(\frac{n+1}{2})!}$.  The product term, with normalisation re-included is then
	\begin{equation}
	\psi_n'(0)\psi_{n+1}(0)=\frac{(n!)^2}{2^{n-1} \sqrt{2 n} (n-1)!\left(\frac{n}{2}!\right)^2},
	\end{equation}
	which through computer algebra software is found to have the limit
	\begin{equation}
	\lim\limits_{n\to \infty}\psi_n'(0)\psi_{n+1}(0) = \frac{2}{\pi}.
	\end{equation}
	Using this in Eq.(\ref{eq:progress}), we find
	\begin{equation}
	q(n)\approx\frac14+\Re\sum_{k=-\infty}^{\infty}e^{-ik\tau}\frac{1}{\pi^2 k^2}.
	\end{equation}
	We can now identify the sum as the exponential Fourier series for a symmetric triangle wave, with amplitude $\frac14$, which is exactly the classical correlator.  This represents the classicalization of the QHO, as $n$ becomes large. 
	\subsection{QHO quasi-probability with arbitrary coarse-grainings}
	\label{app:arbproj} 
    To calculate the quasi-probability in the QHO with arbitrary coarse-grainings, we apply the method in Appendix \ref{app:general}, using the integrals $J_{k\ell}(a,b)$.
    
	We first note Eq.(\ref{eqn:wronski}) is undefined for $k=l$, and must be calculated by hand as 
	\begin{equation}
	J_{kk}(a,b)=\int_{a}^{b}\mathop{dx}\psi_k(x)\psi_k(x).
	\end{equation}
	The quasi-probability is then calculated through Eq.(\ref{eqn:qpapprox}) as
	\begin{equation}
	q_{nn}(+,+)=e^{i\frac{E_n \tau}{\hbar}}\sum_{k=0}^{\infty}e^{-i\frac{E_k \tau}{\hbar}}J_{nk}(a,b)J_{nk}(c,d).
	\end{equation}
	We note again, that these $\psi_k(x)$ and $E_k$ are not yet the QHO eigenstates and energies, and that this result is generic. 
	
	To make things concrete, we will now calculate this for the ground state of the QHO. We take the first pair of intervals $\Delta(\alpha_1)$ to be $\Delta(+)=[0,\infty)$ and $\Delta(-)=(-\infty, 0]$.  The second pair of intervals $\Delta(\alpha_2)$ are taken to be $\Delta(+)=[c,d]$, and $\Delta(-)$ its complement. The quasi-probability for the ground state is,
	\begin{equation}
	\label{eq:qmult}
	q(+,+)=\frac12 J_{00}(c,d)+\Re\sum_{k=0}^{\infty}e^{-i \omega \tau k}J_{0k}(0,\infty)J_{0k}(c,d).
	\end{equation}  
	Again, due to it being a symmetric potential, we have $J_{00}(0,\infty)=\frac12$ as before.  Doing the integral manually for the ground state, we find
	\begin{equation}
	J_{00}(c,d)=\frac12(\erf(d)-\erf(c)).
	\end{equation}
	Using this result and the QHO eigenspectrum in Eq.(\ref{eqn:wronskiapp}), we are able to calculate the quasi-probability using Eq.(\ref{eq:qmult}).
	\section{Morse Potential}
	\label{app:morse}
		\begin{figure}[b]
		\subfloat[]{{\includegraphics[height=4.9cm]{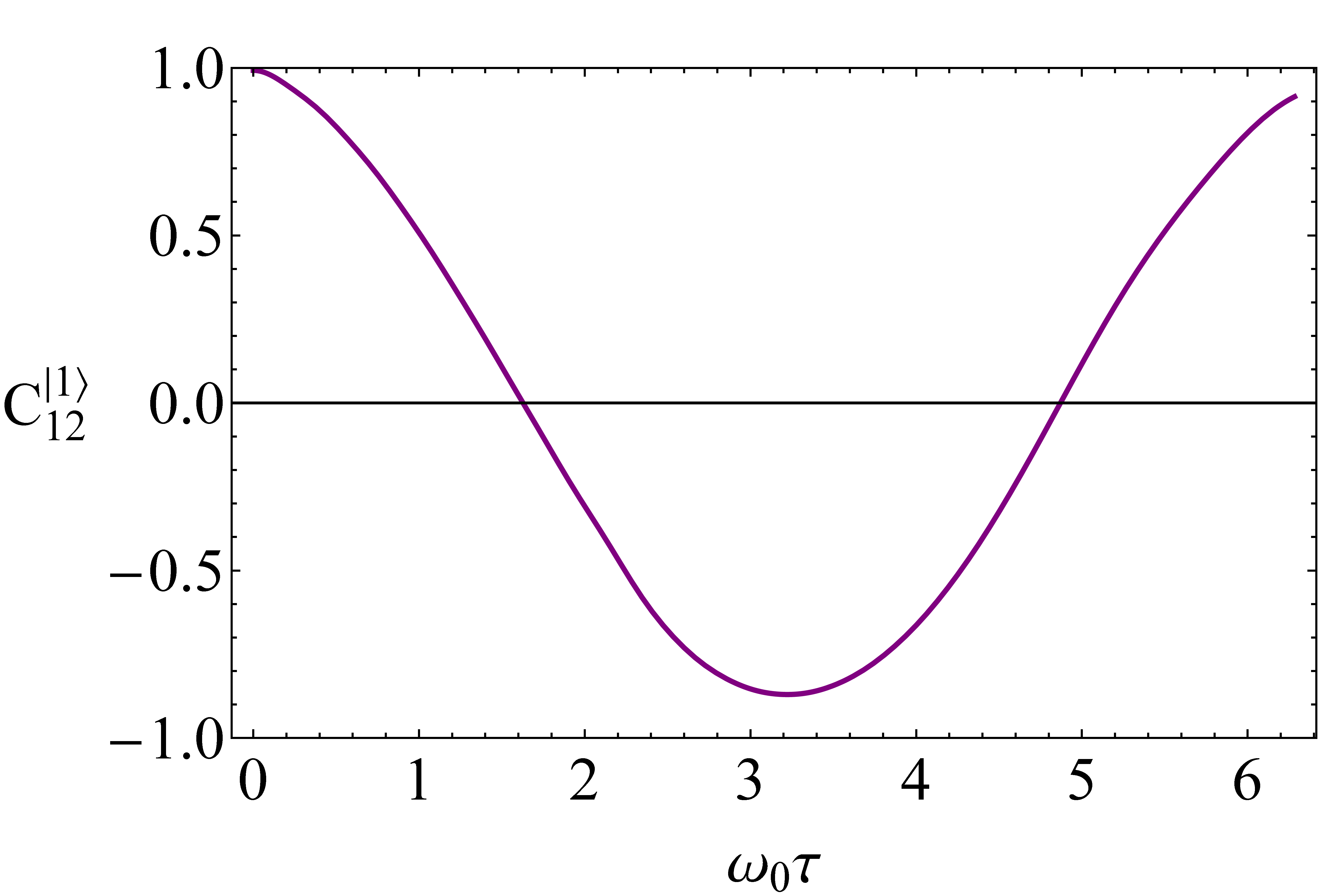}}}%
		\qquad
		\subfloat[]{{\includegraphics[height=4.9cm]{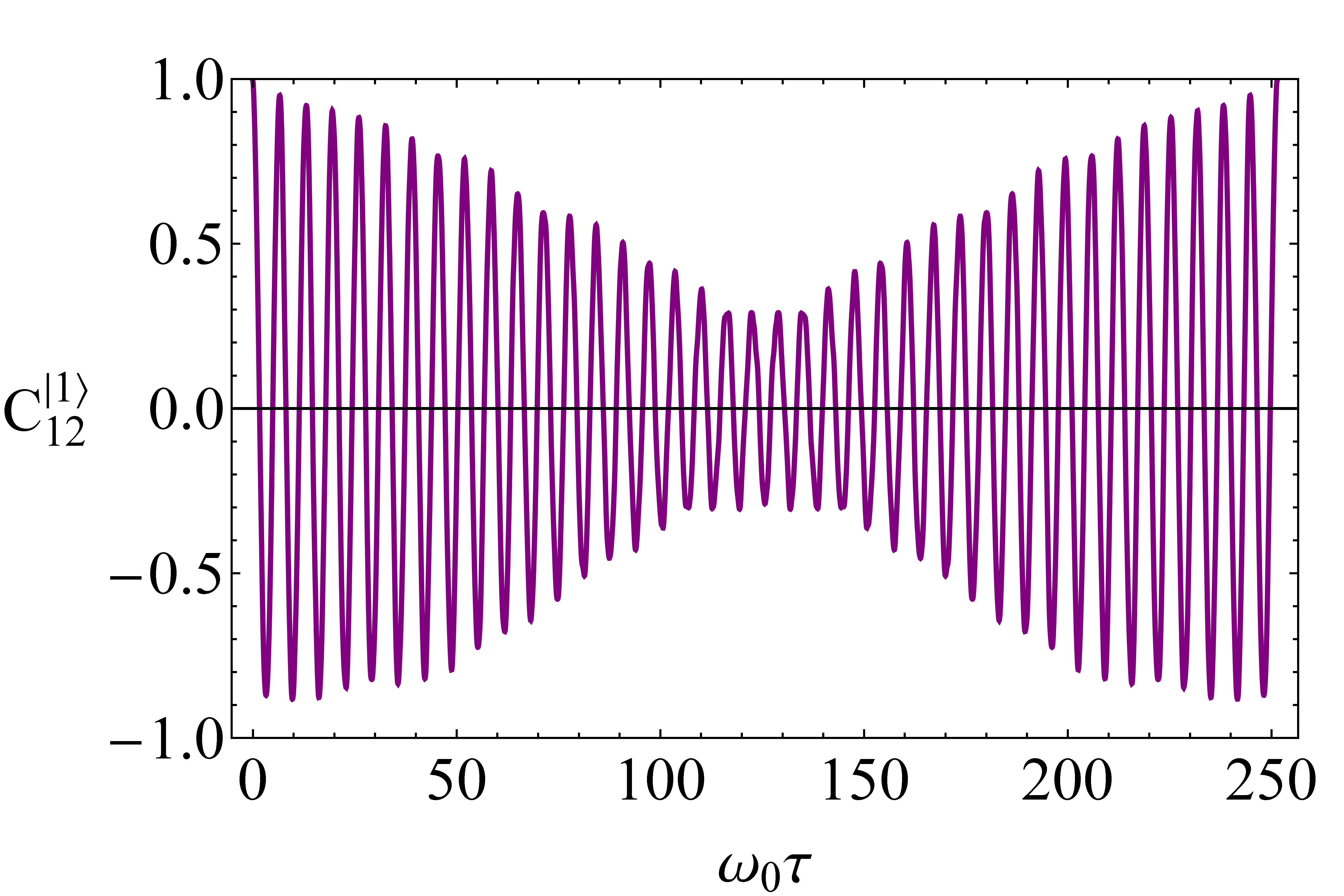}}}
		\caption{
			The temporal correlator in the $\ket1$ state for the Morse potential over time interval $\omega_0 \tau = 2\pi$ is shown in (a). It is not quite periodic over this time interval but becomes exactly periodic over a much longer period of time, as shown in (b).
		}%
		\label{fig:morse}%
	\end{figure}
	The Morse potential is an asymmetric potential with minimum at $r_c$, which combines a short-range repulsion, with a long-range attraction.  The potential is defined by
	\begin{equation}
	V(r) = D_e\left(e^{-2a(r-r_e)}-2e^{-a(r-r_e)}\right),
	\end{equation}
	where $D_e$ corresponds to the well depth, and $a$ to its width.  The Morse potential supports up to $\lfloor \lambda-\frac12 \rfloor$ bound states \cite{dahl1988}, where $\lambda=\frac{\sqrt{2mD_e}}{a\hbar}$ and $\lfloor x \rfloor$ is the floor function.  The eigenstates and energies are
	\begin{align}
	\varepsilon_n&=-\frac12\left(\lambda-n-\frac12\right)^2,\\
	\psi_n(z)&=N_n z^{\lambda-n-\frac12}e^{-\frac12 z}L_n^{2\lambda-n-1}(z),
	\end{align}
	where $L_n^{(\alpha)}(z)$ are the generalized Laguerre polynomials, and $z$ is a scaled spatial coordinate defined as $z=2\lambda e^{-a(r-r_c)}$.
	In the standard non-dimensionalization, physical energy eigenvalues relate to dimensionless as $E_n=\frac{2\hbar \omega_0}{\lambda}\varepsilon_n$, 	where $\omega_0=a \sqrt{\frac{2D_e}{m}}$ is the frequency of small oscillations in the potential.  Normalisation is given by
	\begin{equation}
	N_n=\left(\frac{n!(2\lambda-2n-1)}{\Gamma(2\lambda-n)}\right).
	\end{equation}
	In the Morse potential, states are constrained to $r>0$, however we can still consider coarse-graining onto the left and right halves of the well.  The energy eigenstates similarly still vanish at $r=0$ and $r\to\infty$, so many of the terms in Eq.(\ref{eqn:wronski}) still vanish.  Filling out Eq.(\ref{eqn:qpapprox}), we find
	\begin{equation}
	\label{eqn:morseq}
	q_{nn}=J_{nn}^2+e^{i\frac{\varepsilon_n}{\lambda}\omega_0\tau}\sum_{k=0,k\neq n}^{\lfloor \lambda-\frac12 \rfloor}e^{-i \frac{\varepsilon_k}{\lambda}\omega_0\tau}\frac{1}{(\varepsilon_n-\varepsilon_k)^2}\left(\psi_n^\prime(x_c)\psi_k(x_c)-\psi_k^\prime(x_c)\psi_n(x_c)\right)^2,
	\end{equation}	
	Although this sum is finite, by choosing a large enough $\lambda$, and just looking at low energy states we can approximate it well.  Physically, this corresponds to measurements in shallower wells being likely to eject the particle from the well, where we would then have to consider the continuous positive energy solutions as well.  We can again estimate the error for a particular $n$ and $\lambda$ as
	\begin{equation}
	\Delta(n,\lambda)=J_{nn}-\sum_{k=0}^{\lfloor \lambda-\frac12 \rfloor} J_{nk}^2.
	\end{equation}
		\begin{figure}
		\subfloat[]{{\includegraphics[height=5.2cm]{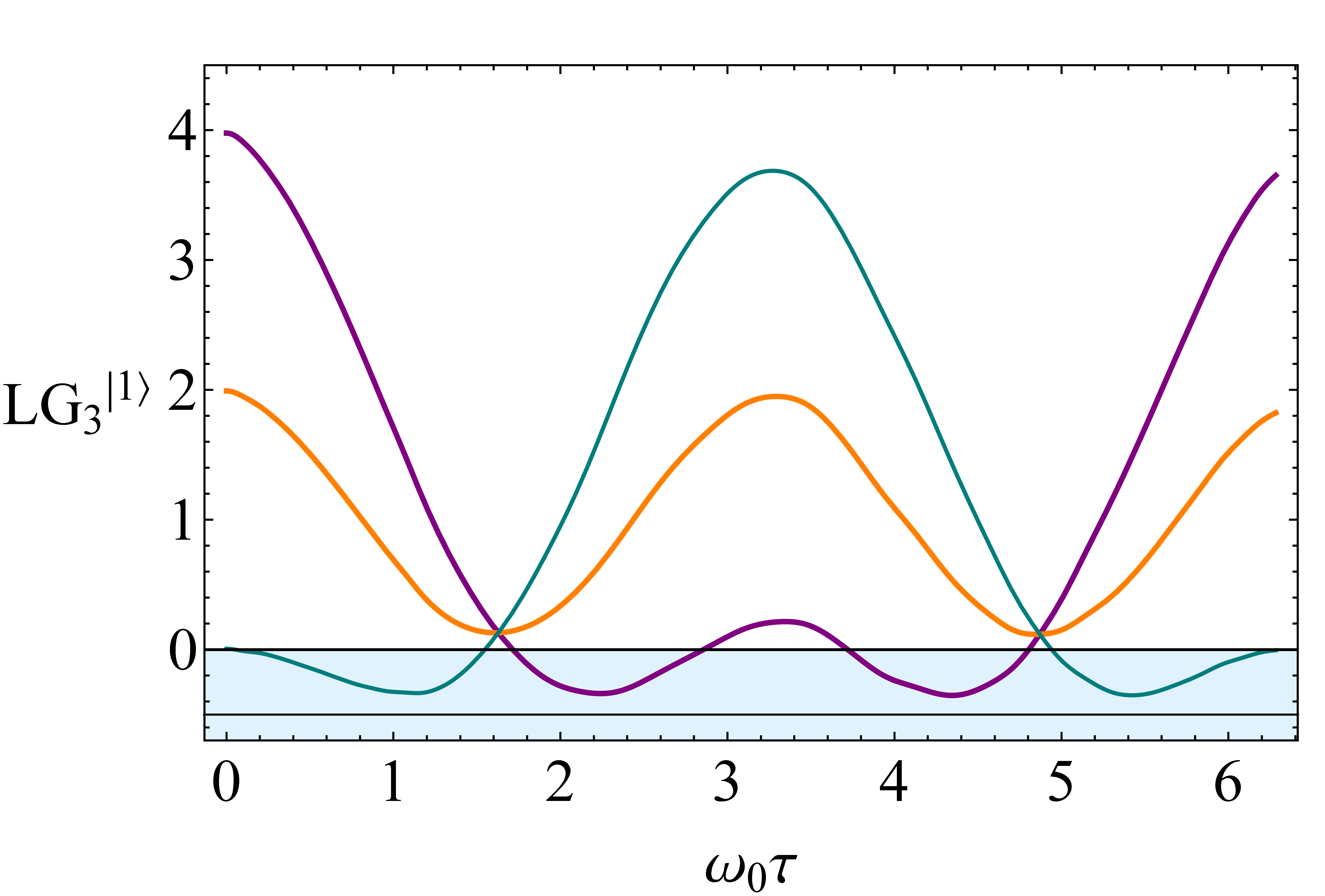}}}%
		\qquad
		\subfloat[]{{\includegraphics[height=5.2cm]{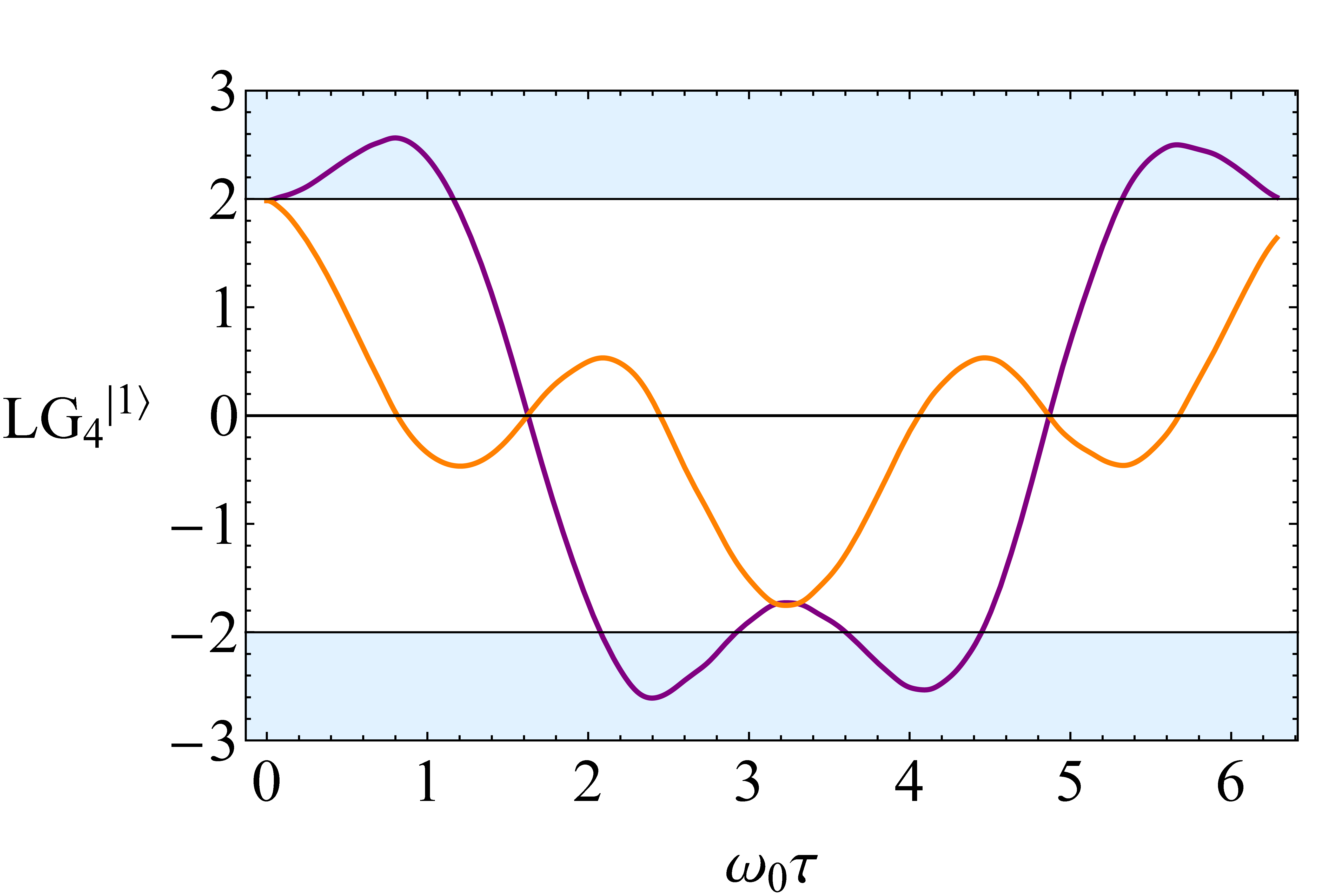}}}
		\caption{In (a), the LG3 inequalities for the $\ket1$ state of the Morse potential with $\lambda=50$ is plotted.  In (b), the LG4 inequalities are plotted.  There are significant violations for both, reaching $70\%$ and $92\%$ of their maximal violations respectively.}%
		\label{fig:morse2}%
	\end{figure}
	Owing to the fact the Morse potential is non-symmetric, we must calculate the terms $J_{nn}$ by hand, which for the low $n$ is a simple integration.  The truncation error for the first excited state reaches $0.01$ for $\lambda=15$, and similar to the QHO, is remains higher for the ground state.  For accuracy, we hence calculate just the $\ket1$ state, and choose $\lambda=50$, yielding $\Delta(n,\lambda)=0.001$.
	
	We extract temporal correlators from Eq.(\ref{eqn:morseq}) using the moment expansion Eq.(\ref{momexp}), which we plot in Fig.~\ref{fig:morse}(a).  At first glance, this correlator is qualitatively very similar to the $\ket1$ correlator for the QHO, and hence similar to the simple cosine correlator for spin-$\frac12$ models.  However owing to the asymmetry of the Morse potential, the correlator never reaches the value $-1$,  and due to the anharmonicity, the correlator is periodic over a much longer time scale, which is shown in Fig.~\ref{fig:morse}(b).
	
	In Fig.~\ref{fig:morse2}(a), we plot the LG3 inequalities for the first excited state of the Morse potential, where there are significant violations, reaching $70\%$ of the L\"{u}ders bound.  The LG4 inequalities are plotted in Fig.~\ref{fig:morse2}(b), with significant violations reaching $92\%$ of the L\"{u}ders bound.  Violations diminish in magnitude for the large-time behaviour, but remain present for nearly all intervals of $\omega_0 \tau$.

	\bibliography{bib}

\end{document}